\documentclass[a4paper,oneside,10pt]{amsart}

\usepackage{amsmath,amssymb,epsf}
\usepackage{amsfonts,amsbsy,amscd,stmaryrd}
\usepackage{latexsym,amsopn}
\usepackage{amsthm}
\usepackage{esint}
\usepackage{mathrsfs}
\allowdisplaybreaks
 
\usepackage{graphicx}

\newcounter{smallarabics}
\newenvironment{arabicenumerate}
{\begin{list}{{\normalfont\textrm{(\arabic{smallarabics})}}}
  {\usecounter{smallarabics}\setlength{\itemindent}{0cm}
   \setlength{\leftmargin}{5ex}\setlength{\labelwidth}{4ex}
   \setlength{\topsep}{0.75\parsep}\setlength{\partopsep}{0ex}
   \setlength{\itemsep}{0ex}}}
{\end{list}}

\newcounter{smallroman}

\newcommand{\ben}{\begin{arabicenumerate}}  
\newcommand{\een}{\end{arabicenumerate}}

\def\init{\setcounter{equation}{0}}

\newtheorem{theorem}{Theorem}[section]

\newtheorem{proposition}[theorem]{Proposition}
\newtheorem{lemma}[theorem]{Lemma}

\theoremstyle{definition}
\newtheorem{definition}[theorem]{Definition}
\newtheorem{remark}[theorem]{Remark}
\newtheorem{example}[theorem]{Example}
\newcommand{\beq}{\begin{equation}}
\newcommand{\eeq}{\end{equation}}

\newcommand{\bea}{\begin{aligned}}
\newcommand{\eea}{\end{aligned}}

\newcommand{\bex}{\begin{example}}
\newcommand{\eex}{\end{example}}
\def\bel{\begin{lemma}}
\def\eel{\end{lemma}}
\def\bet{\begin{theoreme}}
\def\eet{\end{theoreme}}
\def\bed{\begin{definition}}
\def\eed{\end{definition}}
\def\ber{\begin{remark}}
\def\eer{\end{remark}}

\def\rr{{\mathbb R}}
\def\zz{{\mathbb Z}}

\def\nn{{\mathbb N}}

\def\ss{{\mathbb S}}

\def\part{{\rm par}}

\def\H{{H}}

\def\bar{\overline}

\def\proof{
\noindent{\bf Proof.}\ \ }

\usepackage{calrsfs}
\DeclareMathAlphabet{\pazocal}{OMS}{zplm}{m}{n}

\def\cD{{\pazocal D}}
\def\cU{{\pazocal U}}

\def\cM{{\pazocal M}}

\def\cN{{\pazocal N}}
\def\cO{{\pazocal O}}
\def\cC{{\pazocal C}}
\def\cW{{\pazocal W}}

\def\wf{{\rm WF}}

\let\dotlessi\i
\def\i{{\rm i}}

\def\loc{{\rm loc}}

\let\Im\relax

\DeclareMathOperator{\Im}{Im}

\DeclareMathOperator{\Dom}{Dom}

\newcommand{\qeds}{\qed\medskip}

\def \p{ \partial}

\def\12{\frac{1}{2}}
\def\14{\frac{1}{4}}

\def\e{{\rm e}}

\DeclareMathOperator{\supp}{supp}

\newcommand{\one}{\boldsymbol{1}}

\def\cH{{\pazocal H}}

\def\c{{\rm c}}
\def\F{{\rm F\,}}
\def\aF{\rm \bar{F}}

\def\cC{{\pazocal C}}

\def\12{\frac{1}{2}}

\def\e{{\rm e}}

\def\Diff{{\rm Diff}}

\def\bep{\begin{proposition}}
\def\eep{\end{proposition}}
\def\b{{\rm b}}

\def\cE{\pazocal{E}}

\def\init{\setcounter{equation}{0}}

\DeclareSymbolFont{boldoperators}{OT1}{cmr}{bx}{n}
\SetSymbolFont{boldoperators}{bold}{OT1}{cmr}{bx}{n}

\makeatletter
\newcommand*{\defeq}{\mathrel{\rlap{%
                     \raisebox{0.3ex}{$\m@th\cdot$}}%
                     \raisebox{-0.3ex}{$\m@th\cdot$}}%
                     =}
                     
\makeatother
\makeatletter
\newcommand*{\eqdef}{=\mathrel{\rlap{%
                     \raisebox{0.3ex}{$\m@th\cdot$}}%
                     \raisebox{-0.3ex}{$\m@th\cdot$}}%
                     }
\makeatother

\def\Sol{{\rm Sol}}

\def\Op{{\rm Op}}
\def\WF{{\rm WF}}

\newcommand{\traa}[1]{\mskip-6mu\upharpoonright_{#1}}
\newcommand{\tra}[1]{\mskip-10mu\upharpoonright_{#1}}

\def\cf{\pazocal{C}^\infty}
\def\cfd{\dot{\pazocal{C}}^\infty}

\def\zero{{\rm\textit{o}}}
\def\be{{}^{\rm b}}
\def\c{{\rm c}}

\def\inti{{\circ}}

\def\tdiag{t\text{-}{\rm diag}}
\def\cWb{\cW_\b^{-\infty}(X)}

\def\loc{{\rm loc}}
\newcommand{\Hl}{H_{0,\b,\loc}}

\newcommand{\Hb}{H_{0,\b}}
\newcommand{\Hc}{H_{0,\b,{\rm c}}}

\newcommand{\Hp}{H_{0,\b,+}}
\newcommand{\Hm}{H_{0,\b,-}}
\newcommand{\Hpm}{H_{0,\b,\pm}}

\let\origmaketitle\maketitle
\def\maketitle{
  \begingroup
  \def\uppercasenonmath##1{} 
  \let\MakeUppercase\relax 
	\origmaketitle
  \endgroup
}

\newcommand{\AdS}{{\rm AdS}}
\newcommand{\GBB}{{\rm GBB}}
\newcommand{\pt}{({\rm PT})}
\newcommand{\tf}{({\rm TF})}
\newcommand{\inv}{{\scriptscriptstyle (-1)}}
\def\dN{\dot{\pazocal N}}

\newcommand{\dv}{{([0,\epsilon)_x;\cD'(\p X))}}

\newcommand{\bH}[1]{H^{0,#1}_{0,\b,\rm loc}(X)}

\def\pX{\partial X}
\def\deco{^{1,\infty}_{0,\b}}

\def\pX{\partial X}

\def\mass{\nu^2 - \textstyle\frac{(n-1)^2}{4}} 

\begin{document}

\title[The holographic Hadamard condition on asymptotically $\AdS$ spacetimes]{\Large The holographic Hadamard condition\\ on asymptotically Anti-de Sitter spacetimes}

\author{\normalsize Micha{\l} \textsc{Wrochna}}
\address{Universit\'e Grenoble Alpes, Institut Fourier, UMR 5582 CNRS, CS 40700, 38058 Grenoble \textsc{Cedex} 09, France}
\email{michal.wrochna@univ-grenoble-alpes.fr}
\keywords{Quantum Field Theory on curved spacetimes, asymptotically Anti-de Sitter spacetimes, holography, Hadamard condition}
\subjclass[2010]{81T13, 81T20, 35S05, 35S35}

\begin{abstract}In the setting of asymptotically Anti-de Sitter spacetimes, we consider Klein-Gordon fields subject to Dirichlet boundary conditions, with mass satisfying the Breitenlohner--Freedman bound. We introduce a condition on the $\b$-wave front set of two-point functions of quantum fields, which locally in the bulk amounts to the usual Hadamard condition, and which moreover allows to estimate wave front sets for the holographically induced theory on the boundary. We prove the existence of two-point functions satisfying this condition, and show their uniqueness modulo terms that have smooth Schwartz kernel in the bulk and have smooth restriction to the boundary. Finally, using Vasy's propagation of singularities theorem, we prove an analogue of Duistermaat \& H\"ormander's theorem on distinguished parametrices.
\end{abstract}

\maketitle

\section{Introduction and summary of results}

\subsection{Introduction}

The mathematically rigorous formulation of Quantum Field Theory on globally hyperbolic spacetimes, established throughout the last few decades and comprehensively summarized in a handful of recent reviews \cite{HW,benini,FR15,FV2,KM,FR16}, crucially relies on the overcoming of difficulties caused by the generic absence of symmetries. A particularly important step was the replacement of the Killing symmetry-based concept of vacuum state  by a class of physical states satisfying the so-called Hadamard condition \cite{FSW,FNW,KW,radzikowski}, and the implementation of this idea into the perturbative construction of interacting theories \cite{BF00}. The study of \emph{Hadamard states} is now an active field of research, to mention only a couple of recent works on constructive and conceptual aspects and applications \cite{FV,BJ,brumfredenhagen,sanders,VW,FMR,GOW}. 

The assumption that the spacetime is globally hyperbolic narrows however the range of applications, as this excludes for instance Anti-de Sitter space (widely studied in the context of the $\AdS$/${\rm CFT}$ correspondence \cite{maldacena}), even though many symmetry-based  constructions were successfully developed in that particular case, see e.g.~\cite{isham,rehren2,DR1,DR2,BEM02,ishibashiwald,DR3,KW15,BFQ,DF}.

The goal of the present paper is the rigorous construction of non-interacting scalar quantum fields  on \emph{asymptotically} $\AdS$ spacetimes, assuming Dirichlet boundary conditions at the horizon. We use the algebraic approach and propose what we call the \emph{holographic Hadamard condition}. We prove that states satisfying this condition exist indeed and their two-point functions are unique modulo terms that are smooth in the bulk. Moreover, as we will see, a similar statement holds true for the induced conformal field theory on the boundary.

\subsection*{Classical fields} Before discussing our results in more detail let us give an overview of results in the setting of asymptotically $\AdS$ and related spacetimes, starting with classical fields.

The Klein-Gordon equation on Anti-de Sitter was studied by Breitenlohner and Freedman \cite{breiten}, who showed its solvability in a certain mass regime (cf.~the work of Yagdjian and Galstian, who found an explicit solution \cite{yagdjian}), and by Ishibashi and Wald \cite{ishibashiwald}, who described the static dynamics corresponding to different boundary conditions.  An analogous result to that of \cite{breiten} for the Dirac equation was obtained by Bachelot \cite{bachelot3}. Solvability with Dirichlet boundary conditions in the more general case of asymptotically AdS spacetimes was established by Holzegel \cite{holzegel} and reworked by Vasy \cite{vasy}, who proved propagation of singularities theorems. Neumann and Robin boundary conditions were investigated by Warnick \cite{warnick1}, and a study of other boundary conditions was recently performed by Holzegel, Luk, Smulevici and Warnick \cite{hlsw}, see also the related work of Bachelot \cite{bachelot2} in the $\AdS$ case, and of Gannot \cite{gannot} in the static case. Applications to holography were studied by Enciso and Kamran and the higher form Proca equation was studied in the general framework of conformal geometry by Gover, Latini and Waldron \cite{GLW}. The Klein-Gordon and Dirac equation on asymptotically AdS black hole spacetimes are the subject of many recent developments, including \cite{holzegelwarnick,HS,guillaume,warnick2,gannot2,dold}.

\subsection*{Quantum fields} Quantum Field Theory on $\AdS$ spacetime was studied by Avis, Isham and Storey \cite{isham}, who based their analysis on exact formulae for  solutions (and bi-solutions) of the Klein-Gordon equation in terms of hypergeometric functions. This approach was further developed in a rigorous language by Bros, Epstein and Moschella in \cite{BEM02}.

The widespread interest in the foundations and the consequences of the $\AdS$/${\rm CFT}$ correspondence \cite{maldacena} raised questions on how the $\AdS$/${\rm CFT}$ duality can be transferred to the ground of quantum fields, and what are its manifestations on the level of  observables. This was clarified by the works of Rehren \cite{rehren,rehren2} (who proposed what is now known as Rehren duality, cf.~\cite{ribeiro1,ribeiro2} for generalizations to asymptotically $\AdS$ spacetimes) and D\"utsch and Rehren \cite{DR1,DR2,DR3}, and was further studied by Kay and Larkin \cite{KL} and Kay and Ort{\'\dotlessi}z \cite{KO}. We also refer to the recent work of Zahn \cite{zahn} for a holographic prescription with features similar to the field-theoretical $\AdS$/${\rm CFT}$, though in a different setting. 

Wald \cite{wald1} and Ishibashi and Wald \cite{ishibashiwald0,ishibashiwald} laid ground for the construction of quantum fields on static asymptotically $\AdS$ by studying the classical static dynamics and clarifying the r\^ole of different boundary conditions, though the full analysis was only performed on $\AdS$. Useful related results for the Poincar\'e patch of $\AdS$, including Strichartz estimates, were obtained by Bachelot \cite{bachelot4} (cf.~\cite{bachelot} for a more refined analysis focused on de Sitter branes).

Advances on globally hyperbolic spacetimes based on the Hadamard condition (in particular its applications in renormalization) have triggered studies of the local behaviour of two-point functions on $\AdS$ and other non-globally hyperbolic spacetimes, using the Hadamard parametrix as main ingredient \cite{Kay,KW15,BFQ,DF}. So far however no `microlocal' proposal in the spirit of Radzikowski's fundamental work \cite{radzikowski} has been made (though formal computations involving a wave front set condition are already present in Morrison's work on $\AdS$ \cite{morrison}), and it is unclear how to incorporate non-static spacetimes or holography in present local approaches. 

\subsection{Setup} The point of view adopted in the present paper is that while on a globally hyperbolic spacetime, singularities of solutions of the Klein-Gordon equation (and hence of two-point functions) are naturally described using the wave front set, on asymptotically $\AdS$ spacetimes it is useful to use the \emph{$\b$-wave front set}, as motivated by Vasy's propagation of singularities theorem \cite{vasy}. 

Let us first introduce the setup very briefly. An asymptotically $\AdS$ spacetime is a manifold $X$ with boundary (denoted $\pX$) equipped with a Lorentzian metric $g$, which near $\pX$ is of the form
\[
g=\frac{-dx^2+h}{x^2}
\]  
for some symmetric two-tensor $h$ with Lorentzian restriction to $\pX$ (see Definition \ref{defads} for the precise formulation). We consider the  Klein-Gordon operator on $(X,g)$,
\[
P\defeq\Box_g + \mass, 
\]
where $n\geq 2$ is the dimension of $X$, and we assume $\nu>0$ (this is the so-called Breitenlohner--Freedman bound \cite{breiten}).

One of the outcomes of \cite{vasy} is the existence of Dirichlet retarded and advanced propagators $P_\pm^{-1}$, i.e., inverses of $P$ that solve $Pu=f$ and $u\traa{\pX}=0$ for $u$ and $f$ vanishing at respectively past and future infinity. We show that  there is a natural space of solutions denoted by $\Sol_{0,\b}^{1,\infty}(P)$, which  in view of  mapping properties of $P_\pm^{-1}$ can be characterized as the range of the following isomorphism:
\[
P_+^{-1}-P_-^{-1}: \ \frac{ \Hc^{-1,\infty}(X) }{P  \Hc^{1,\infty}(X)}\longrightarrow\Sol\deco(P).
\] 
Here, $\Hc^{1,\infty}(X)$ (resp.~$\Hc^{-1,\infty}(X)$) is the space of compactly supported distributions, \emph{conormal} with respect to the zero Sobolev space $H_0^1(X)$ (resp.~to $H^{-1}_0(X)$, the dual of $H_0^1(X)$). These spaces are defined in \eqref{ss:conormal}, for the moment we only state their most essential features: elements of $\Hc^{1,\infty}(X)$ are smooth in the interior $X^\inti$, belong to $L^2(X,g)$, and possess extra regularity with respect to vector fields tangent to the boundary (that is, `conormal regularity' or `$\b$-regularity', as opposed to `smooth regularity' relative to all vector fields), furthermore the `$\c$' subscript indicates that the support is contained in a compact time interval.  

Solutions in $\Sol\deco(P)$ are locally in $\Hc^{1,\infty}(X)$, and for that reason we regard them as being `maximally regular'. More generally, if $u$ is a distribution (and if it belongs to the dual of $\Hc^{-1,\infty}(X)$), one introduces a set $\wf_\b^{1,\infty}(u)$ (the $\b$-wave front set of $u$) which indicates where microlocally $u$ fails to be in $\Hc^{1,\infty}(X)$. Vasy's theorem describes then the propagation of $\wf_\b^{1,\infty}(u)$ given some information about $\b$-regularity of $Pu$ \cite{vasy}. Though in the interior of $X$, this locally amounts to H\"ormander's propagation of singularities theorem, the additional feature is that singularities are reflected upon reaching the boundary.

\subsection{Main results} In this setting, analogies with the globally hyperbolic case lead us to consider two-point functions to be pairs of operators $\Lambda^\pm$ that satisfy:
\[
\bea i) \quad & P\Lambda^\pm = \Lambda^\pm P =0,\\
ii) \quad & \Lambda^+-\Lambda^- = \i (P_+^{-1}- P_{-}^{-1}) \,\mbox{ and }\, \Lambda^\pm\geq 0.
\eea
\]
We say that $\Lambda^\pm$ satisfy the \emph{holographic Hadamard condition} if
\beq\label{hhad}
\wf'_\b(\Lambda^\pm)\subset \dot\cN^\pm\times\dot\cN^\pm,
\eeq 
where $\wf'_\b$ is an operatorial version of the $\b$-wave front set (which is defined in Subsect.~\ref{ss:WFb}, and which is different from the operatorial $\b$-wave front set often considered in the literature on $\b$-calculus, although closely related), and $\dot\cN^\pm$ are the positive/negative energy components of the compressed bicharacteristic set $\dot\cN$ of $P$. To explain it very briefly, let us first denote by $\tilde g$ the `desingularized' conformally rescaled metric $x^2 g$. The  \emph{compressed bicharacteristic set} $\dot\cN$ is obtained from the characteristic set $\cN$ of $\Box_{\tilde g}$ by identifying covectors with the same tangential momentum but different normal momenta at $\pX$. Thus, condition \eqref{hhad} is practically the same as the Hadamard condition on globally hyperbolic spacetimes (in the formulation of \cite{SV,hollands}, which is equivalent to Radzikowski's original one \cite{radzikowski}), with the main difference being the possibility that singularities are reflected at the horizon. Indeed we show that \eqref{hhad} implies a more specific form of $\wf'_\b(\Lambda^\pm)$ that captures this phenomenon. 

Our main result can be stated as follows.

\begin{theorem}[See Thm.~\ref{thm:existence} \& Prop. \ref{cor1}] Two-point functions $\Lambda^\pm$ satisfying the holographic Hadamard condition \eqref{hhad} exist and are unique modulo terms whose Schwartz kernel is smooth in the interior $X^\inti$.
\end{theorem}
 
The existence is proved using an adaptation of the deformation argument of Fulling, Narcowich and Wald \cite{FNW}, originally proposed for globally hyperbolic spacetimes.\medskip

Using Vasy's propagation of singularity theorem we also prove an analogue of Duistermaat \& H\"ormander's theorem \cite{DH72} on distinguished parametrices (strictly speaking formulated here in terms of inverses) in the present setting. Namely, we show that there are four inverses of $P$ which are uniquely determined modulo regularizing terms (in the sense of $\b$-regularity) by their primed $\b$-wave front sets, see Theorem \ref{prop:wfs} for the full statement.

The crucial ingredient underpinning these results and Vasy's work is Melrose's $\b$-calculus \cite{melrose88,melrose}, see Appendix \ref{appb} for a brief introduction. It is worth mentioning that this formalism has been succesfully applied to General Relativity and plays an important r\^ole in the recently announced resolution of the Kerr-de Sitter stability conjecture by Hintz and Vasy \cite{HV}. It was also recently applied to Quantum Field Theory (on asymptotically Minkowski spacetimes) \cite{GHV,positive,VW}, though in the present work it is used in a different way.   

\medskip

In our terminology, the word \emph{holographic} refers to additional features of two-point functions satisfying \eqref{hhad}. To explain this, let us first recall some basic aspects of the field theoretical $\AdS$--$\rm CFT$ correspondence (see \cite{rehren} for a more detailed introduction), here in the more general setup of asymptotically $\AdS$ spacetimes. A brief inspection of the equation $P u=0$ leads one to expect that the solutions are of the form
\beq\label{eq:slgksdfg}
u = x^{\nu_+} v_+ + x^{\nu_-} v_-, \ \ \nu_\pm =\frac{n-1}{2}\pm \nu,
\eeq
with $v_-=0$ in our case since Dirichlet boundary conditions are imposed. Extending an argument due to Vasy \cite{desitter,vasy} we show \eqref{eq:slgksdfg} to be true for $u$ conormal in $x$ with values in distributions on $\pX$, and moreover, we show that this implies $v_+\in\cf\dv$. This means in particular that the weighted restriction
\[
\p_+ u = (x^{-\nu_+}u)\traa{\pX}
\] 
is well defined. Since $x^{-\nu_+}u=v_+$ is smooth in the direction normal to the boundary, the information about conormal regularity of $u$ given by $\wf_\b^{1,\infty}(u)$ can be used to estimate the (usual, `smooth') wave front set of $\p_+ u$. 

The field theoretical $\AdS$--$\rm CFT$ correspondence sets to promote the operation $\p_+$ to the level of quantum fields, and thus in terms of two-point functions, the relevant object to study is $\p_+ \Lambda^\pm \p_+^*$. We prove:

\begin{theorem}[See Thm.~\ref{hololo}]  If $\Lambda^\pm$ are two-point functions satisfying the holographic Hadamard condition \eqref{hhad}, then $\wf'(\p_+ \Lambda^\pm \p_+^*)\subset\pm (\Gamma\times\Gamma)$ for some $\Gamma\subset T^*\pX\setminus\zero$ with $\Gamma\cap-\Gamma=\emptyset$ (where the minus sign means multiplication by $-1$ in the covariables). Furthermore, if $\tilde\Lambda^\pm$ is another such pair of two-point functions then $\p_+(\tilde\Lambda^\pm-\Lambda^\pm )\p_+^*$ have smooth Schwartz kernel. 
\end{theorem}

Using the terminology of generalized free fields on curved spacetimes introduced in \cite{sanders2}, $\Lambda^\pm$ induce boundary-to-boundary two-point functions $\p_+ \Lambda^\pm \p_+^*$ that satisfy the so-called generalized Hadamard condition (see Thm.~\ref{hololo} for a more detailed description of the set $\Gamma$ in the present case). This is in agreement with what one expects basing on known properties of generalized free fields on the boundary of $\AdS$, see e.g.~\cite{DR2}. 

\subsection{Outlook} The main question that arises from our results is whether the formalism of perturbative algebraic QFT \cite{BF00,HW1,HW2,FR15,dang} can be adapted to construct interacting theories on asymptotically $\AdS$ spacetimes and to relate them with ${\rm CFT}$s on the boundary.

It would also be desirable to have a more direct construction of holographic Hadamard states, for instance in the spirit of the works \cite{junker,GW,GOW}.

Another open issue are boundary conditions other that Dirichlet ones: useful hints are provided by the recent work of Dappiaggi and Ferreira (which considers a local Hadamard condition in the bulk) \cite{DF}, as well as the works \cite{bachelot2,warnick1,gannot,hlsw,kamran} which deal with classical fields. We conjecture that in the case of Neumann and Robin boundary conditions, a condition similar to our holographic Hadamard condition \eqref{hhad} can be consistently formulated, with similar consequences for holography, though it is likely that this will have to involve conormality with respect to a different space than the one considered here (i.e. the zero-Sobolev space $H^1_0(X)$, see the main part of the text); some advances along those lines can be found in \cite{gannot}.

\subsection{Plan of the paper} In Section \ref{sec:KG} we introduce the geometrical setup and we recall results due to Vasy which are essential to our analysis.

In Section \ref{sec:sol} we construct the symplectic space of conormal solutions of the Klein-Gordon equation with Dirichlet boundary conditions and prove several auxiliary results on holography. 

Section \ref{eq:ref} discusses the particular case of static asymptotically $\AdS$ spacetimes, in which case the classical evolution of $P$ is shown to be directly related to a model equation of the form $\p_t^2 + A$, with $A$ a (positive) self-adjoint operator on a Hilbert space. 

In Section \ref{sec:propagators} we introduce the operatorial $\b$-wave front set $\wf'_\b$ and discuss its basic properties. We then define two-point functions in the present setting and introduce the holographic Hadamard condition. We prove the existence by reduction to the static case using a deformation argument. We then give an analogue of Duistermaat \& H\"ormander's theorem, as outlined in the introduction, and study weighted restrictions of holographic Hadamard two-point functions.

Appendix \ref{appb} contains a brief introduction to the calculus of $\b$-pseudo\-differential operators used throughout the paper.

\section{The Klein-Gordon equation on asymptotically $\AdS$ spacetimes}\init\label{sec:KG}

\subsection{Notation} If $X$ is a smooth manifold with boundary $\p X$, we denote by $X^\inti$ its interior. We denote by $\cf(X)$ the space of smooth functions on $X$ (in the sense of extendability across the boundary). The space of smooth functions vanishing with all derivatives  at the boundary $\p X$ are denoted by $\dot\cC^{\infty}(X)$, and their dual by $\cC^{-\infty}(X)$. Their compactly supported counterparts are denoted respectively by $\cf_\c(X)$, $\dot\cC^{\infty}_\c(X)$, $\cC^{-\infty}_\c(X)$. 

On the boundaryless manifold $\p X$ we use the conventional notation $\cD'(\pX)$ for the space of distributions and $\cE'(\pX)$ for compactly supported ones. 

The signature of Lorentzian metrics is taken to be $(+,-,\dots,-)$. Furthermore, we adopt the convention that sesquilinear forms $(\cdot|\cdot)$ are linear in the second argument. 

\subsection{Asymptotically $\AdS$ spacetimes} The spacetime of interest is modelled by an $n$-di\-men\-sio\-nal ($n\geq 2$) smooth manifold $X$ with boundary $\p X$ (also called in this context the \emph{horizon}), and its interior $X^\inti$ is equipped with a Lorentzian metric $g$. Let $x$ be a boundary-defining function of $\p X$. We recall at this point that given $x$, there exists $W\supseteq\pX$, $\epsilon>0$ and a diffeomorphism $\phi:[0,\epsilon)\times \pX \to W$ such that $x\circ \phi$ agrees with the projection to the first component of $[0,\epsilon)\times \pX$. We always assume that such $\phi$ is already given and drop it in the notation subsequently.

 We employ Vasy's definition of asymptotically $\AdS$ spacetimes \cite{vasy}:

\begin{definition}\label{defads} $(X,g)$ is called an asymptotically Anti-de Sitter (\AdS) spacetime  if near $\p X$, the metric $g$ is of the form
\beq\label{eq:formproduct}
g=\frac{-dx^2+h}{x^2},
\eeq
with $h\in\cf(X;{\rm Sym}^2 T^* X)$ such that with respect to some product decomposition $X=\pX\times [0,\epsilon)_x$ near $\p X$, the restriction $h\traa{\pX}$ is a section of $T^*\pX\otimes T^*\pX$ and is a Lorentzian metric on $\pX$.  
\end{definition}

We refer the interested reader to \cite[Def.~6, Lem.~2.3]{gannot} for a discussion of sufficient conditions that give a metric of the form \eqref{eq:formproduct}, cf.~\cite{CG} for remarks on how asymptotically $\AdS$ spacetimes fit into the general framework of conformal geometry. We remark that in the literature, often more restrictive definitions are considered, see e.g.~\cite{holzegel,warnick1,kamran}.

We denote by $\tilde g$ the conformally related metric
\[
\tilde g\defeq x^2 g,
\]
and so $\tilde g=-dx^2+h$ near $\p X$. Definition \ref{defads} implies that $\p X$ is time-like with respect to $\tilde g$, meaning that the dual metric $\tilde g^{-1}$ of $\tilde g$ is negative definite on $N^* \p X$, the conormal bundle of $\p X$ in $X$ (or put differently, $\tilde g^{-1}(dx,dx)<0$ at $\p X$).

\subsubsection{Universal cover of $\AdS$} The basic example of an asymptotically $\AdS$ spacetime is the universal cover $(X_{\AdS},g_{\AdS})$ of Anti-de Sitter space (the universal cover is needed to rule out closed time-like curves, which would spoil the global results we are interested in, see e.g.~\cite{isham,vasy}). Its interior $X_\AdS^\inti$ is modelled by $\rr\times \rr^{n-1}$ and the metric there is given by
\[
g_{\AdS}=(1+r^2) dt ^2 -(1+r^2)^{-1} dr^2   - r^2 d\omega^2,
\]
expressed here in coordinates $(t,r,\omega)$ (commonly simply referred to as `$\AdS$ spherical coordinates'), valid away from $r=0$, where $\omega$ are the standard coordinates on the sphere. The change of coordinates $x=r^{-1}$ allows one to compactify $\rr^{n-1}$ to a ball $\mathbb{B}^{n-1}$ and to include a boundary, $\p X_{\AdS}=\{x=0\}$, so that $X_{\AdS}= \rr_t\times \overline{\mathbb{B}^{n-1}}$. This way, a collar neighborhood of $\pX_{\AdS}$ can be identified with $\rr_t\times [0,1)_x \times \ss_\omega^{n-2}$, and the metric becomes
\[
g_{\AdS}=\frac{(1+x^2)dt^2 -(1+x^2)^{-1}dx^2 - d\omega^2}{x^2}
\]
in that neighborhood, which is of the form required in Definition \ref{defads}.

\subsection{Klein-Gordon equation and $\b$-geometry}\label{ss:b} Our main object of interest will be the Klein-Gordon operator\footnote{Recall the convention $(+,-,\dots,-)$ for the Lorentzian signature. Throughout the paper, $\Box_g=\frac{1}{\sqrt{|g|}}\p_\mu(\sqrt{|g|} g^{\mu\nu}\p_\nu)$.}
\[
P\defeq\Box_g + \mass, \ \  \nu>0.
\]
on an asymptotically $\AdS$ spacetime $(X,g)$. In what follows we recall the notions needed for the geometrical description of the propagation of the singularities of its solutions. Recall that in the interior $X^\inti$, the bicharacteristics of $P$ are the integral curves of the Hamilton vector field $\H_p$ of the principal symbol $p$ restricted to the characteristic set $\cN = p^{-1}(\{0\})$, see e.g.~\cite{hoermander}.
As $\tilde g$ is conformally related to $g$, one can equally well use the principal symbol $\tilde p$ of $\Box_{\tilde g}$ to define the characteristic set $\cN$ and the bicharacteristics. Since $\tilde g$ is smooth down to the boundary, it makes thus sense to redefine
\[
\cN=\tilde{p}^{-1}(\{ 0\})\subset T^* X.
\]
Turning our attention to issues arising at the boundary, we adopt the point of view that the propagation of singularities of solutions of $P$ is best described as taking place in the \emph{$\b$-cotangent bundle}, as advocated by Melrose and worked out in the present setting by Vasy. We briefly recall the relevant definitions, working in local coordinates $(x,y)=(x,y_1,\dots,y_{n_1})$ on $X$, where $x$ is as usual a boundary defining function of $\p X$. The starting point is the observation that smooth vector fields that are tangent to the boundary are in the $\cf(X)$-span of $x\p_x$ and $\p_{y_i}$, $i=1,\dots,n-1$, and thus can be viewed as smooth sections of a vector bundle, denoted $\be TX$. The $\b$-cotangent bundle, $\be T^*X$, is then defined as the dual bundle of $\be TX$. This way, smooth sections of  $\be T^*X$ are in the $\cf(X)$-span of $\frac{dx}{x}$ and $dy_i$, $i=1,\dots,n-1$.

If $U\subset X$ we denote by $T^*_U X$,  $\be T^*_U X$ the restriction over $U$ of the respective bundles. 

Writing $\xi,\zeta$ for the covariables relative to $x,y$, there is a natural map $\varpi: T^*X\to \be T^*X$ which in our coordinates is given by 
\beq
\varpi(x,y,\xi,\zeta)= (x,y,x\xi,\zeta).
\eeq
Away from $\p X$, $\varpi$ is a diffeomorphism that allows one to identify $T^*_{X^\inti}X$ with $\be T^*_{X^\inti}X$. On the other hand, over $\p X$ the map $\varpi$ is no longer one-to-one; it defines however a useful \textit{embedding} of $T^*\pX$ into $\be T^*_{\pX} X$.

The \emph{compressed characteristic set} of $P$ is 
\[
\dN\defeq \varpi(\cN)\subset \be T^*X.
\]

We use Vasy's definition of generalized broken bicharacteristics, which is primarily based on earlier work by Lebeau \cite{lebeau}. 

\begin{definition}\label{def:gbb} A \emph{generalized broken bicharacteristic} of $P$ (or, in short, a $\GBB$) is a continuous map $\gamma:I\to \dN$ defined on an interval $I\subset \rr$, satisfying:
\beq
\liminf_{s\to s_0}\frac{(f\circ\gamma)(s)-(f\circ\gamma)(s_0)}{s-s_0}\geq \inf \big\{ \H_{\tilde p}(\varpi^*f)(q): \ q\in \varpi^{-1}(\gamma(s_0))\cap\cN\big\}
\eeq
for all $f\in\cf(\be T^*X)$.
\end{definition}

Over the interior $X^\inti$, $\varpi$ is one-to-one and thus Definition \ref{def:gbb} means that in $X^\inti$, $\gamma$ is made of integral curves of the Hamilton vector field of $p$. In the general case, Definition \ref{def:gbb} accounts for the possibility that $f\circ \gamma$ is not differentiable, which happens as a consequence of $\varpi$ not being one-to-one.

Crucially, let us stress that in Definition \ref{def:gbb}, $\gamma$ is required to be continuous as a map $I\to\be T^*X$ (thus, `in $x\xi$' rather than `in $\xi$') and so the normal momentum is allowed to jump. In other words, $\GBB$s can be reflected at the boundary.  We refer to \cite{vasy,corners,edges} for a more detailed description of $\GBB$s, cf.~\cite{MVW} for the more intricate setup of edge manifolds.

\subsection{Conormal regularity}\label{ss:conormal}

One of the essential features in Vasy's approach to the Klein-Gordon equation on $\AdS$ is the interplay between the class of \textit{$\b$-differential operators} $\Diff_\b(X)$, defined as the algebra generated by smooth vector fields tangent to the boundary, and the algebra of \emph{$0$-differential operators} $\Diff_0(X)$, generated by smooth vector fields vanishing on the boundary. Using local coordinates $(x,y)$ near $\p X$, the former, $\Diff_\b(X)$, is $\cf(X)$-generated by $x\p_x$ and $\p_{y_i}$, $i=1,\dots,n$. It is essential for studying conormal regularity. On the other hand, $\Diff_0(X)$ is $\cf(X)$-generated by $x\p_x$ and $x\p_{y_i}$; this `degenerate' subclass of $\Diff_\b(X)$ arises naturally as we have $P\in\Diff_0(X)$ in the present setup. 

We denote by $(\cdot|\cdot)_{L^2}$ the inner product of $L^2(X)=L^2(X,g)$. Sometimes we will also use the $L^2(X,\tilde g)$ inner product for the rescaled metric $\tilde g = x^2 g$; note the relation $L^2(X)= x^{\frac{n}{2}}L^2(X,\tilde g)$. Recall that if $Q\in \Diff(X)$ then its formal adjoint $Q^*\in \Diff(X)$ is defined by $(\phi_1 | Q \phi_2)_{L^2}=( Q^* \phi_1 |  \phi_2)_{L^2}$ for all $\phi_1,\phi_2\in\cfd_\c(X)$. One important property of $\b$-differential operators is that if $Q\in \Diff_\b(X)$ then the identity $(\phi_1 | Q \phi_2)_{L^2}=( Q^* \phi_1 |  \phi_2)_{L^2}$ extends to all $\phi_1,\phi_2\in \cf_\c(X)$, i.e. there are no boundary terms.  

We will work in the setting of Sobolev spaces $H^{k,s}_{0,\b}(X)$, which distinguish between regularity with respect to $\Diff_0(X)$ and $\Diff_\b(X)$. First, if $k$ is a non-negative integer $k$, one defines
\[
 H^k_0(X)=\big\{  u\in \cC^{-\infty}(X): \ Qu\in L^2(X)\ \forall Q\in \Diff_0^k(X)  \big\},  
\]
where the superscript $k$ in $\Diff_0^k(X)$ refers to the differential operator's order (in the very usual sense) and we recall that $L^2(X)$ is defined using the volume form of $g$. This space is topologized using the norm
\[
\| u \|_{H^k_0}=\| u \|_{L^2} +  \sum_i \| Q_i u \|_{L^2},
\]
where $\{Q_i\}_{i=1,\dots,N}$ is an arbitrarily chosen collection of elements of $\Diff_0^k(X)$ such that at each point, at least one $Q_i$ is elliptic, see \cite{MM}. The definition generalizes to negative integers e.g.~by letting $H^{-k}_0(X)$ be the dual of  $H^{k}_0(X)$ (relative to the $L^2(X)$ pairing). Then, if $s\geq 0$ is an integer, $H^{k,s}_{0,\b}(X)$ is by definition
\beq\label{eq:defks}
 H^{k,s}_{0,\b}(X)=\big\{  u\in H_0^k(X): \ Qu\in H_0^k(X)\ \forall Q\in\Diff_\b^s(X)  \big\}, 
\eeq
with norm 
\[
\| u \|_{ H^{k,s}_{0,\b}}=\| u \|_{H^k_0} + \sum_i \| Q_i u \|_{H^k_0},
\]
where $\{Q_i\}_{i=1,\dots,N}$ is an arbitrarily chosen collection of elements of $\Diff_\b^s (X)$ such that at each point, at least one $Q_i$ is elliptic (see Appendix \ref{appb}). The definition can be extended to negative integers in such way that $H^{-k,-s}_{0,\b}(X)$ is the dual space of $H^{k,s}_{0,\b}(X)$. We remark here that in the interior, say for compact $K\subset X^\inti$, $H^{k,s}_{0,\b}(K)$ is just $H^{k+s}(K)$, whereas at the boundary, $H^{k,s}_{0,\b}(X)$ distinguishes between `$0$-regularity' and `$\b$-regularity'.
 
One denotes by $\Hc^{k,s}(X)$ the subspace of compactly supported elements of $\Hb^{k,s}(X)$, and by  $\Hl^{k,s}(X)$ the space of all $u\in\cC^{-\infty}(X)$ such that $\chi u \in H^{k,s}_{0,\b}(X)$ for all $\chi\in \cf_\c(X)$. The spaces $\Hc^{k,s}(X)$ and $\Hl^{k,s}(X)$ are topologized in the usual way. Namely, $\Hc^{k,s}(X)$ is equipped with the strongest locally convex topology such that for all compact $K\subset X$, the embedding of $\Hb^{k,s}(K)$ (the space of all $u\in\Hb^{k,s}(X)$ supported in $K$) into $\Hc^{k,s}(X)$ is continuous. Furthermore, the topology of $\Hl^{k,s}(X)$ is given by the seminorms $\|u\|_{H^{k,s}_{0,\b},\chi}=\|\chi u\|_{H^{k,s}_{0,\b}}$, where $\chi$ runs over $\cC_\c^\infty(X)$. The important feature of these topologies is that a map
\[
\Lambda : \Hc^{k_1,s_1}(X)\to \Hl^{k_2,s_2}(X)
\]
is continuous if and only if $\chi\Lambda :  H_{0,\b}^{k_1,s_1}(K)\to H_{0,\b}^{k_2,s_2}(X)$ is continuous for all $K\subset X$ compact and $\chi\in\cf_\c(X)$, where $\chi\Lambda$ acts on $\Hc^{k_1,s_1}(K)$ via the embedding of  $H_{0,\b}^{k_1,s_1}(K)$ in $\Hc^{k_1,s_1}(X)$.

Finally, we let
\[
\Hb^{k,\infty}(X)\defeq\textstyle\bigcap_s H^{k,s}_{0,\b}(X), \ \ \Hb^{k,-\infty}(X)\defeq\textstyle\bigcup_s H^{k,s}_{0,\b}(X),
\]
equipped with their canonical Fr\'echet space topologies, and similarly as before we define the spaces $\Hc^{k,\pm\infty}(X)$, $\Hl^{k,\pm\infty}(X)$ correspondingly. 

\subsection{Retarded/advanced problem and propagation of singularities theorems}\label{ss:wb}

Let us recall that our main object of interest is the Klein-Gordon operator
\[
P=\Box_g + \mass, \quad \nu>0, 
\]
on an asymptotically $\AdS$ spacetime $(X,g)$, with Dirichlet boundary conditions at $\pX$. The assumption $\nu>0$ will be made throughout the whole paper.

In what follows we recall results due to Vasy \cite{vasy} which will be the starting point in our analysis. 

Let us denote by $\pi:\be T^*X \to X$ the bundle projection. Following \cite{vasy}, we make the following two global assumptions:
\begin{enumerate}
\item[$(\rm TF)$] there exists $t\in \cf(X)$ such that for every GBB $\gamma$, $t\circ\pi\circ \gamma : \rr\times \rr$ is either strictly increasing or strictly decreasing and has range $\rr$;\vspace{0.1cm}
\item[$(\rm PT)$] topologically, $X=\rr_t\times\Sigma$ for some compact manifold $\Sigma$ with boundary.
\end{enumerate}

From $(\rm PT)$ it follows that the map $t:X\to\rr$ is proper, which is the condition assumed originally in \cite{vasy}. We remark that the universal cover of $\AdS$ satisfies the two conditions $(\rm TF)$, $(\rm PT)$.

\begin{theorem}[{\cite[Thm.~1.6]{vasy}}]\label{thm:vasy1} Assume the two hypotheses $\tf$ and $\pt$. Let $t_0,s\in\rr$, $s'\leq s$. Suppose
\beq
f\in \Hl^{-1,s+1}(X), \ \ \supp f \subset \{t\geq t_0\}.
\eeq
Then there exists a unique $u\in \Hl^{1,s'}(X)$ that solves the retarded problem
\beq
Pu=f,\ \ \supp u \subset \{t\geq t_0\}.
\eeq
Furthermore, $u$ is in fact in $\Hl^{1,s}(X)$, and for all compact $K\subset X$ there exists a compact $K'\subset X$ and a constant $C>0$ such that
\[
\| u \|_{\Hb^{1,s}(K)}\leq C \| f \|_{\Hb^{-1,s+1}(K')}.
\]
\end{theorem}

The analogous statement for the advanced problem holds true as well.

Note that in Theorem \ref{thm:vasy1}, the Dirichlet boundary conditions are implicitly assumed via the choice of function spaces (this essentially amounts to $\cfd(X)$ being dense in $\Hl^{-1,\infty}(X)$, and can be seen more explicitly by considering asymptotics of solutions; see \cite{vasy} or Section \ref{sec:sol} for more details).\medskip

We will also need microlocal elliptic regularity and propagation of singularities theorems, with singularities being characterized by the \emph{$\b$-wave front set} relative to $H^{k}_{0}(X)$. To define the latter one needs pseudodifferential operator classes $\Psi^s_\b(X)$ (more precisely, `classical' ones) that generalize the $\b$-differential operators $\Diff^s_\b(X)$ of order $s$. These are introduced in Appendix \ref{appb}. Here, without going into details, we just recall that any $A\in \Psi^s_\b(X)$ has a \emph{principal symbol} $\sigma_{\b,s}(A)$, which is a function on $\be T^*X \setminus \zero$.  Now if $k$ is an integer\footnote{All relevant definitions can be easily extended to non-integer $k$, though.} and $u\in H_{0,\b}^{k,-\infty}(X)$, one says that $q\in \be T^*X \setminus \zero$ is \emph{not} in $\wf_\b^{k,\infty}(u)$ if there exists $A\in\Psi^0_\b(X)$ such that  $\sigma_{\b,s}(A)$ is invertible at $q$ and $Au\in \Hl^{k,\infty}(X)$. With this definition, in the interior of $X$, $\wf_\b^{k,\infty}(u)$ is just the usual wave front set, i.e.
\[
\wf_\b^{k,\infty}(u) \cap T^*X^\inti=\wf(u)
\]
using the embedding of $T^* X^\inti$ in $T^*_{X^\inti} X$, which is in turn identified with $\be T^*_{X^\inti} X$.   
Generally over $X$, $\wf_\b^{k,\infty}(u)$ contains information about where microlocally $u$ is not conormal (with respect to $H^{k}_{0}(X))$.

Vasy's propagation of singularities result can be stated as follows (note that it uses neither the $\tf$ hy\-po\-the\-sis nor $\pt$).

\begin{theorem}[{\cite[Thm.~1.5]{vasy}}] Suppose $u\in\Hl^{1,k}(X)$ for some $k\in\rr$. Then
\[
\WF_\b^{1,\infty}(u)\setminus\dN\subset\wf_\b^{-1,\infty}(Pu).
\]
Moreover, the set
\[
\big(\WF_\b^{1,\infty}(u)\cap\dN\big)\setminus \wf_\b^{-1,\infty}(Pu)
\]
is a union of maximally extended $\GBB$s in $\dN\setminus \WF_\b^{-1,\infty}(Pu)$. In particular, if $Pu=0$ then $\WF_\b^{1,\infty}(u)\subset\dN$ is a union of maximally extended $\GBB$s. 
\end{theorem}
 
Thus, singularities of solutions of $Pu=0$ propagate along $\GBB$s; in particular they are reflected at the horizon.

\section{Symplectic space of solutions and holography}\label{sec:sol}

\subsection{Symplectic space of solutions}

Let us denote by $\Hpm^{k,\infty}(X)$ the space of future/past supported elements of $\Hl^{k,\infty}(X)$, i.e.
\beq\label{eq:dsdfsd}
\Hpm^{k,\infty}(X)= \big\{ u\in \Hl^{k,\infty}(X): \  \supp u \subset \{\pm t\geq \pm t_0\} \mbox{ for some } t_0\in\rr\big\}.  
\eeq
Observe that by hypothesis $\pt$ and the above definition, the intersection of those spaces satisfies
\beq\label{eqinter}
\Hp^{k,\infty}(X)\cap\Hm^{k,\infty}(X)\subset \Hc^{k,\infty}(X),
\eeq
where we recall that the additional subscript in $\Hc^{k,\infty}(X)$ refers to the support being compact (note that in the present setup this means support in a compact time interval).

Theorem \ref{thm:vasy1} entails the existence of \emph{Dirichlet retarded/advanced propagators}, denoted respectively $P_\pm^{-1}$, which we consider in the present context to be the unique operators
\beq\label{pp0}
P_\pm^{-1}: \Hpm^{-1,\infty}(X)\to \Hpm^{1,\infty}(X)
\eeq
that satisfy
\beq\label{pp1}
\bea
 P P_\pm^{-1} & = \one \,\mbox{ on }\, \Hpm^{-1,\infty}(X),\\
P_\pm^{-1} P & = \one \,\mbox{ on }\, \Hpm^{1,\infty}(X).
\eea
\eeq
Continuity properties of $P_\pm^{-1}$ can be read off from the exact statement of Theorem \ref{thm:vasy1}, which also implies that $P_\pm^{-1}$ extends uniquely to a map
\[
P_\pm^{-1}: \Hpm^{-1,-\infty}(X)\to \Hpm^{1,-\infty}(X),
\]
where $\Hpm^{k,-\infty}(X)$ is defined in analogy to \eqref{eq:dsdfsd} with $-\infty$ instead of $+\infty$. 

The difference of the two propagators,
\beq\label{eq:defG}
G\defeq P_+^{-1}-P_-^{-1} : \Hc^{-1,\infty}(X)\to \Hl^{1,\infty}(X),
\eeq
will be called the \emph{(Dirichlet) causal propagator}, in agreement with the terminology commonly used on globally hyperbolic spacetimes (one also uses the name \emph{Pauli-Jordan} or \emph{commutator function}). A natural space of solutions is given by
\[
\Sol\deco(P)\defeq \{ u \in \Hl^{1,\infty}(X) : \ Pu =0\}.
\]
We show that this space can be obtained as the range of $G$ on a suitable space, and moreover, $G$ can be used to construct a `symplectic form' on $\Sol\deco(P)$. 

\begin{proposition}\label{prop:symp} The causal propagator \eqref{eq:defG} induces a bijection
\beq\label{eq:iso1}
[G]:\frac{ \Hc^{-1,\infty}(X) }{P  \Hc^{1,\infty}(X)}\longrightarrow\Sol\deco(P).
\eeq
Moreover, $\i (\cdot|  G\cdot)_{L^2}$ induces a non-degenerate hermitian form on the quotient space  $\Hc^{-1,\infty}(X) /P  \Hc^{1,\infty}(X)$.
\end{proposition}
\proof To prove that \eqref{eq:iso1} is well defined, one needs to check that $G \Hc^{-1,\infty}(X) \subset \Sol\deco(P)$ and that $GP=0$ on $\Hc^{-1,\infty}(X)$; both properties follow directly from the relevant definitions.

Injectivity of \eqref{eq:iso1} means that if $f\in \Hc^{-1,\infty}(X)$ and $Gf=0$, then $f=Pu$ for some $u\in \Hc^{1,\infty}(X)$. Indeed, if we set $u=P^{-1}_+f$ then $u\in \Hp^{1,\infty}(X)$ and $Pu=f$. Since $Gf=0$, $u$ can also be written as $u=P^{-1}_- f\in\Hm^{1,\infty}(X)$, and so belongs to $\Hm^{1,\infty}(X)\cap\Hp^{1,\infty}(X)$. In view of \eqref{eqinter}, $u\in \Hc^{1,\infty}(X)$ as requested.

We now turn our attention to surjectivity  of \eqref{eq:iso1}. Let $\chi_\pm\in\cf(X)\cap \Hpm^{1,\infty}(X)$ (that is, $\chi_\pm$ is a future/past supported smooth function) such that $\chi_+ + \chi_-=1$. Then any $u\in\Sol\deco(P)$ can be written as
\beq\label{eq:ernfl}
\bea
u&=\chi_+ u + \chi_- u=P_+^{-1}P\chi_+ u + P_-^{-1}P\chi_- u\\
 &=P_+^{-1}P\chi_+ u + P_-^{-1}P(1-\chi_+) u \\
&= P_+^{-1}P\chi_+ u - P_-^{-1}P \chi_+ u = G P\chi_+ u.  
\eea
\eeq
Since $P\chi_+ u=-P\chi_- u\in \Hm^{1,\infty}(X)\cap\Hp^{1,\infty}(X)\subset \Hc^{1,\infty}(X)$, the computation above shows            that $u=Gw$ for some $w\in\Hc^{1,\infty}(X)$; this gives surjectivity of $[G]$.

For the last claim we need to show that $(P_+^{-1})^*=P_-^{-1}$ as sesquilinear forms on $\Hc^{-1,\infty}(X)$ (well-definiteness of the sesquilinear form induced by $G$ and its non-degeneracy are then easy to conclude). If $f,h\in\Hc^{-1,\infty}(X)$, we have
\[
( f | P_+^{-1} h )_{L^2}=( P  P^{-1}_- f | P_+^{-1} h )_{L^2}
=(P^{-1}_- f|P P_+^{-1}h)_{L^2}=(P^{-1}_- f|h)_{L^2},
\]
where in the second equality we have used that $P$ is formally self-adjoint, belongs to $\Diff_\b(X)$ (so there are no terms supported in $\pX$), and $\supp P^{-1}_- f\cap \supp P_+^{-1} h$ is compact. This proves the assertion.\qeds

We have shown in \eqref{eq:ernfl} that if $\chi\in\cf(X)$ is future supported and $\one-\chi$ is past supported, then 
\beq\label{eq:gpchi}
G[P,\chi]=\one \,\mbox{ on }\, \Sol\deco(X).
\eeq
For any $t_1\neq t_2$, if we choose $\chi$ that equals $1$ in a neighbourhood of $[t_2,\infty)$ and $0$ in a neighbourhood  of $(-\infty,t_1]$, then $[P,\chi]$ vanishes on a neighbourhood of $\rr\setminus[t_1,t_2]$. This means that in the isomorphism \eqref{eq:iso1} we can replace $\Hc^{-1,\infty}(X)$ by $H^{-1,\infty}_{0,\b,[t_1,t_2]}(X)$, the space of all $f\in \Hc^{-1,\infty}(X)$ supported in the region of $X$ in which $t\in [t_1,t_2]$.  In consequence, one obtains from \eqref{eq:gpchi} and Proposition \ref{prop:symp}  the \emph{time-slice property} (or time-slice axiom), which can be formulated as follows. 

\begin{proposition}\label{tslice} The inclusion map $\imath_{t_1,t_2}:H^{-1,\infty}_{0,\b,[t_1,t_2]}(X)  \to \Hc^{-1,\infty}(X)$ induces an isomorphism
\[
[\imath_{t_1,t_2}]:\frac{ H^{-1,\infty}_{0,\b,[t_1,t_2]}(X)  }{P  \Hc^{1,\infty}(X)\cap H^{-1,\infty}_{0,\b,[t_1,t_2]}(X)}\longrightarrow\frac{\Hc^{-1,\infty}(X)}{P  \Hc^{1,\infty}(X)}.
\]
\end{proposition}

In other words, each equivalence class in the quotient space $\Hc^{-1,\infty}/P  \Hc^{1,\infty}$ has a representative that is supported in $[t_1,t_2]$. The field-theoretical interpretation of this is that the full content of the classical field theory can be recovered from data in an arbitrarily small time-interval $[t_1,t_2]$. 

We note that the inverse of $[\imath_{t_1,t_2}]$ is given by $[\imath_{t_1,t_2}^{-1}]$, where
\[
\imath_{t_1,t_2}^{-1} f = [P,\chi] G f = f + ([P,\chi] G f - f).
\]
Indeed, $\imath_{t_1,t_2}^{-1} f$ has the required support properties as $[P,\chi]$ vanishes on a neighbourhood of $\rr\setminus[t_1,t_2]$, and $v=[P,\chi] G f - f$ satisfies $Gv=0$ and so belongs to $P  \Hc^{1,\infty}(X)$.

\begin{remark}One can view Proposition \ref{prop:symp} as the construction of the classical (non-interacting, scalar) field theory on $(X,g)$ associated with Dirichlet boundary conditions. We stress that although on globally hyperbolic spacetimes the standard construction proceeds by considering the space of space-compact solutions of the Klein-Gordon equation (i.e., those with compact intersection with a Cauchy surface, see e.g.~\cite{BGP07}), in the asymptotically $\AdS$ case this is no longer a sensible choice as solutions with initial data supported away from the boundary can reach $\pX$ nevertheless.
\end{remark}

\subsection{Boundary data and holography}\label{ss:holo} We will now be interested in what happens close to the boundary, and so, for the sake of simplicity of notation we will work on $[0,\epsilon)_x \times \pX$.

Let $\nu_\pm=\frac{n-1}{2}\pm \nu$ be the two \emph{indicial roots} of $P$. We assume as in the rest of the paper $\nu>0$. On the other hand, the conditions $(\rm TF)$ and $(\rm PT)$ are unessential for the results in this subsection.

 We will give a distributional version of Vasy's result on asymptotics of (approximate) solutions close to the boundary \cite[Prop. 8.10]{vasy}. The proof is fully analogous to the smooth case considered in \cite{desitter,vasy} (cf.~\cite{gover} for related results in the broad framework of conformally compact manifolds), we repeat it however for the reader's convenience. We start by the construction of approximate solutions from holographic data.

\begin{lemma}\label{lem:correct} Suppose $w\in\cf\dv$ and
\beq\label{eq:tempysd}
P x^\alpha w \in  x^{\alpha+k} \cf\dv
\eeq
for some $k\in\nn_0$, $\alpha>\nu_+ - k$. Then there exists $v\in \cf\dv$ such that
\beq\label{dfsdfe}
P x^\alpha v \in \cfd\dv, \ \ v-w\in x^k \cf\dv.
\eeq
Moreover, if $x^\alpha w\in\Hl^{1,\infty}(X)$ then we can find $v$ as above such that $x^\alpha v\in\Hl^{1,\infty}(X)$. 

The assertions above remain true if $\cD'(\pX)$ is replaced by $\cf(\pX)$.  
\end{lemma}
\proof The crucial property of $P$ that we use is that it can be written as 
\[
P=Q_1 + x Q_2, \ \ Q_1 =(- x\p_x+n-1)x\p_x + \mass, \ \ Q_2\in\Diff_\b^2(X).
\]
One concludes that $P$ acting on distributions of the form $x^\alpha w$ gives
\beq\label{qlkw}
P x^\alpha w = c_\alpha x^\alpha w + x^{\alpha+1} Q_{3,\alpha}w, \ \ Q_{3,\alpha}\in\Diff^2(X),
\eeq
where $c_\alpha=\alpha(n-1)-\alpha^2+\mass$. In particular if $\alpha=\nu_+$ then $c_{\nu_+}=0$ and \eqref{qlkw} simplifies to
\beq\label{qlkw2}
P x^{\nu_+} w = x^{\nu_+ +1} Q_{3,\nu_+}w.
\eeq

The identities (\ref{qlkw}--\ref{qlkw2}) imply that if \eqref{eq:tempysd} holds true, then we can correct $w$ by a term $x^k w_k\in x^k \cf\dv$ to have 
\[
P x^\alpha (w+x^k w_k) \in  x^{\alpha+k+1} \cf\dv,
\]
namely, we set $w_k = -c_{\alpha+k}^{-1}x^{-\alpha-k}Px^\alpha w$. By repeating this step for $k+1,k+2,\dots$ and using Borel summation we obtain $v$ satisfying \eqref{dfsdfe}. Moreover, if $x^\alpha$ belongs to $\Hl^{1,\infty}(X)$ then by construction  all the terms $x^{\alpha+k}w_k$ belong to $\Hl^{1,\infty}(X)$. 

The $\cf(\pX)$ case is proved analogously.\qeds

\begin{proposition} Given any $w_0\in \cD'(\pX)$ there exists $u$ of the form 
\beq
u=x^{\nu_+} v, \ \ v\in \cf\dv,
\eeq
such that $v\traa{\p X}=w_0$ and $Pu\in\cfd\dv$. The same is true with $\cD'(\pX)$ replaced by $\cf(\pX)$. 
\end{proposition}
\proof We abbreviate $\cf\dv$, respectively $\cf(X)$, by $\cf$. We observe that
\[
P x^{\nu_+}(1_x\otimes w_0)\in x^{\nu_+ +1} \cf
\] 
in view of \eqref{qlkw2}. Thus, we can apply Lemma \ref{lem:correct} to $x^{\nu_+}w_0$ starting from $k=1$, which produces $u=x^{\nu_+} v$ with the requested properties.\qeds

To get a converse statement we first need another auxiliary lemma (also analogous to \cite{vasy}).

\begin{lemma}\label{lememem} Suppose $u\in x^{\ell} \bH{s}$ and $Q_1 u = f$ with $f\in x^{\ell+1}  \bH{s-2}$
for some $s\in\rr$, $\ell>-\nu$. If $\nu_+\notin (\ell,\ell+1]$ then 
\[
u\in x^{\ell+1}  \bH{s-2}.
\]
Otherwise, $u=x^{\nu_+}w_0 + u_0$ with $u_0\in x^{\ell+1}\bH{s-2}$ and $w_0\in H_{\loc}^{s-2}(\pX)$,  and the map 
\beq\label{eq:dlfkdf}
x^\ell \bH{s} \ni  u \mapsto w_0 \in H_{\loc}^{s-2}(\pX)
\eeq
is continuous (where $x^\ell \bH{s}$ is topologized in the natural way using the topology of $\bH{s}$).
\end{lemma}
\proof Recall that $Q_1 =(- x\p_x+n-1)x\p_x + \nu_+\nu_-\in\Diff^2_\b(X)$, which is actually an ordinary differential operator in the $x$ variable. The equation $Q_1 u =f$ can be reformulated as
\beq\label{elerkn}
u=\cM^{-1}_{\ell}  q(\sigma)^{-1}  \cM_{\ell} f
\eeq   
where $q(\sigma)= (\sigma-\i(n-1))\sigma + \nu_+\nu_-$ and $\cM_{\ell}$ is the shifted Mellin transform in the $x$ variables, i.e.
\[
(\cM_\ell f)(\sigma)=\int_0^\infty x^{-\i\sigma-\ell}f(x) \frac{dx}{x}, \ \  (\cM^{-1}_\ell v)(x)=\frac{1}{2\pi}\int_{\Im\sigma=-\ell} x^{\i\sigma}v(\sigma)d\sigma.
\]
The poles of the meromorphic function $q(\sigma)^{-1}$ are $\nu_-$ and $\nu_+$, and so $q(\sigma)^{-1}  \cM_{\ell} f$ has a meromorphic continuation with poles at $\nu_-$, $\nu_+$. By shifting the contour in the inverse Mellin transform in \eqref{elerkn} we can replace $\cM^{-1}_{\ell}$ by  $\cM_{\ell+1}^{-1}$, possibly at the cost of adding residues of the form $x^{\nu_\pm} w_0$ with $w_0\in H_\loc^{s-2}(\p X)$. Terms of the form  $x^{\nu_-} w_0$ are however eliminated by the assumption $u\in x^{\ell} \bH{s}$, $\ell>-\nu$.
\qed

\begin{proposition}\label{mimi} Suppose that $u\in\bH{-\infty}$, resp. $u\in\bH{\infty}$,  and 
\beq\label{erfwerfwre}
Pu\in\cfd\dv, \ \ \mbox{resp. } Pu\in\cfd(X).
\eeq
Then $u$ is of the form
\beq\label{formu}
\bea
u=x^{\nu_+} v, \ \ &v\in \cf\dv,  \\ &\mbox{resp. } v\in\cf([0,\epsilon)_x;\cf(\p X)).
\eea
\eeq
Furthermore, the map $u\mapsto v\traa{\pX}$ is continuous (using the $\bH{-\infty}$, resp. the $\bH{\infty}$ topology for $u$ and the $\cD'(\pX)$, resp. $\cf(\pX)$ topology for $v$).
\end{proposition}
\proof We focus on the $\bH{-\infty}$ case. Let us first suppose that $u\in x^k \Hl^{0,\infty}(X)$ for some $k\geq 0$, and that 
\beq\label{eqlwkejfwlrkjg}
Pu\in\cfd\dv.
\eeq
Let us recall that the differential operator $P$ can be written as
\[
P=Q_1 + x Q_2, \ \ Q_1 =(- x\p_x+n-1)x\p_x + \nu_+\nu_-, \ \ Q_2\in\Diff_\b^2(X),
\]
We have $x Q_2 u\in x^{k+1}  \Hl^{0,-\infty}(X)$, which in view of \eqref{eqlwkejfwlrkjg} implies 
\[
Q_1 u \in x^{k+1}  \Hl^{0,-\infty}(X).
\]
We use Lemma \ref{lememem}, which asserts that if $\nu_+\notin (k,k+1]$, one has
\[
u\in x^{k+1}  \Hl^{0,-\infty}(X).
\]
Otherwise, one concludes $u=x^{\nu_+}w_0 + u_0$, where 
\[
w_0\in \cD'(\pX), \ \  u_0\in x^{k+1}\Hl^{0,-\infty}(X).
\]
Since by \eqref{qlkw2}, 
\[
P x^{\nu_+} w_0\in x^{\nu_+ + 1}\cf\dv,
\] 
using Lemma \ref{lem:correct} we obtain $v_0$ s.t. 
\[
w_0-v_0\in x\cf\dv,   \ \ P x^{\nu_+} v_0\in\cfd\dv.
\]
Thus,
\[
\bea
u- x^{\nu_+} v_0 &\in x^{k+1}\Hl^{0,-\infty}(X)+ x^{\nu_+ +1}\cf\dv,\\
&\phantom{\in} \subset x^{k+1}\Hl^{0,-\infty}(X),\\
P(u- x^{\nu_+} v_0)& \in \cfd\dv.
\eea
\]
Therefore, we can iterate the whole argument and prove  this way the existence.

In view of how $v$ is constructed, the continuity of $u\mapsto w_0 = v\traa{\pX}$ is a consequence of the continuity of the map \eqref{eq:dlfkdf}. \qeds

For $u\in x^{\nu_+}\cf\dv$ we denote
\beq
\p_+ u = (x^{-\nu_+} u)\traa{\pX},
\eeq
so that on solutions of $Pu=0$, $\p_+$ coincides with the map $u\mapsto v\traa{\pX}$ from Proposition \ref{mimi}.

We are interested in knowing what is the wave front set of $\p_+ u$ given information about the regularity of $u$. 

\begin{lemma}\label{lem:comm} Let $q\in T^*\pX$ and suppose $B\in \Psi^0_\b(X)$ is elliptic at $\varpi(q)\in \be T^*_{\pX} X$. Then there exists $\tilde B_0\in \Psi^0(\p X)$ elliptic at $q$ and such that $\p_+ B = \tilde B_0 \p_+$ on $x^{\nu_+}\cf\dv$. 
\end{lemma}
\proof Let $\tilde B = x^{-\nu_+} B x^{\nu_+}$. Then $\tilde B\in \Psi^0_\b(X)$ is elliptic at $\varpi(q)$ (in fact, $\sigma_0(\tilde B)=\sigma_0(B)$, see Appendix \ref{appb}). Furthermore,
\[
B u = B x^{\nu_+} v = x^{\nu_+} \tilde B v.
\]
Since $\Psi^0_\b(X)$ preserves $\cf\dv$, $\tilde B v \in \cf\dv$. A standard fact on the $\b$-calculus (see e.g.~\cite{melrose88}) says that there exists  $\tilde B_0\in\Psi^0(\p X)$ elliptic at $q$ such that $\tilde B_0 (w\traa{\pX})=(\tilde B w)\traa{\p X}$. Therefore,
\[
\p_+   B u = (\tilde B v)\traa{\pX} = \tilde B_0 (v\traa{\pX}) = \tilde B_0 \p_+ u,
\]
which finishes the proof. \qeds

If $\Gamma\subset \be T^*X$ we use the short-hand notation $\Gamma\traa{\pX}\subset T^*\pX$ for the intersection $\Gamma\cap T^* {\pX}$ defined by means of the embedding of  $T^*\p X$ in $\be T^*_{\pX} X$. 

\begin{proposition}\label{prop:wf} Suppose $u \in\Hl^{1,-\infty}(X)$ and $Pu=0$. Then
\beq
\wf( \p_+ u )\subset (\wf_\b^{1,\infty}(u))\traa{\pX}.
\eeq
\end{proposition}
\proof Let $\varpi(q)\in \be T^* \p X$ and suppose $\varpi(q)\notin \wf_\b^{1,\infty}(u)$, so that there exists $B\in\Psi^0_\b(X)$ elliptic at $\varpi(q)$ such that $Bu\in \Hb^{1,\infty}(X)$. By Proposition \ref{mimi}, 
\[
u\in x^{\nu_+}\cf\dv.
\] 
Since $B$ preserves $x^{\nu_+}\cf\dv$, $Bu\in x^{\nu_+}\cf\dv$. Thus,
\[
P Bu= (Q_1+xQ_2) Bu \in x^{\nu_+ +1}\cf\dv.     
\]
By Lemma \ref{lem:correct}  there exists $\tilde u\in \Hb^{1,\infty}(X)\cap x^{\nu_+}\cf\dv$ such that 
\[
P\tilde u\in \cfd\dv,  \ \ \p_+ \tilde u = \p_+ B u.
\]
Since $P\tilde u$ also belongs to $\Hb^{-1,\infty}(X)$, we have actually $P\tilde u\in \cfd(X)$. We can thus use Proposition \ref{mimi}  to conclude that $\p_+\tilde u\in \cf(\pX)$. By Lemma \ref{lem:comm}, there exists $B_0$ elliptic at $q$ and such that $\p_+ B = \tilde B_0 \p_+$. Thus,
\[
\tilde B_0 \p_+ u  = \p_+ B  u = \p_+ \tilde u \in \cf(\pX).
\]
This shows that $q\notin\wf(\p_+ u)$.
\qed

\section{The static case}\label{eq:ref}

\subsection{Standard static asymptotically $\AdS$ spacetime} In this section we discuss the special class of \emph{static} asymptotically $\AdS$ spacetimes, on which it is possible to simplify the analysis of the Klein-Gordon equation by using arguments from spectral theory.

Recall that in the setting of manifolds without boundary, in any static spacetime there exist local coordinates $(t,w^i)$ in which the metric $\tilde g$ takes the form 
\beq\label{eq:static}
\tilde g =  \beta dt^2 - \tilde g_{ij} dw^i dw^j,
\eeq
where $t$ is the Killing flow parameter and $\beta$, $\tilde g_{ij}$ are $t$-independent smooth coefficients. 

In the following definition the metric is required to be globally of the form \eqref{eq:static}.

\begin{definition} An $n$-dimensional \emph{standard static spacetime} is a Lorentzian manifold $(X^\inti,\tilde g)$ of the form $X^\inti= \rr \times \Sigma^\inti$, with $\Sigma^\inti$ a  manifold of dimension $n-1$, and such that the metric $\tilde g$ is of the form
\[
\tilde g= \beta dt^2 - \pi^*\tilde h, 
\]
where the \emph{static time coordinate} $t:X^\inti\to\rr$ is the canonical projection onto the first factor, $\pi:X^\inti \to \Sigma^\inti$ is the canonical projection onto the second factor, $\tilde h$ is a Riemmanian metric on $\Sigma^\inti$, and $\beta\in\cf(\Sigma^\inti)$ satisfies $\beta>0$.
\end{definition}

We refer to \cite{sanchezstatic} for a more detailed analysis of standard static spacetimes. 

For the sake of brevity we will drop $\pi^*$ in our notation.

We define below a class of asymptotically $\AdS$ spacetimes which is a subclass of stationary ones, considered e.g.~in \cite{gannot}.

\begin{definition} We say that an asymptotically $\AdS$ spacetime $(X,g)$ of dimension $n$ is \emph{standard static} if:
\begin{enumerate}
\item[$(1)$] $X=\rr\times \Sigma$ with $\Sigma$ an $n-1$ dimensional compact manifold with boundary,
\item[$(2)$] there exists a boundary defining function $x$ of $\p X$ as in Definition \ref{defads}  such that, setting $\tilde g = x^2 g$,  $(X^\inti, \tilde g)$ is a standard static spacetime, and moreover, denoting by $t$ the static time coordinate, $\p_t x =0$. 
\end{enumerate}
\end{definition}

We remark that if $(X,g)$ is standard static then the global assumptions $(\rm TF)$, $(\rm PT)$ introduced in Subsect.~\ref{ss:wb} are automatically satisfied.

Let us discuss further implications of standard staticity. By $(2)$, $x^2 g=-dx^2+h$ with $h=\beta dt^2-k$, where $k\in\cf(\Sigma; {\rm Sym}^2 T^* \Sigma)$ and $\beta\in\cf(\Sigma)$ are smooth down to the boundary (since $h$ is). Thus,  $k\traa{\p\Sigma}$ is a Riemannian metric and one also concludes immediately that  $(\pX,h\traa{\p X})$ is a standard static Lorentzian spacetime.

Note that $x\traa{\Sigma}$ (the restriction of $x$ to any time slice) defines a boundary-defining function for $\p \Sigma$. We will write $x$ instead of $x\traa{\Sigma}$ whenever there is no risk of confusion.

Proceeding exactly as in \cite[Sec.~5]{grahamlee} and \cite[Sec.~2.1]{gannot} we can show that near $\p X$, $g$ is of the form
\beq\label{eq:newform}
g=\frac{-dx^2+\beta(x) dt^2-k(x)}{x^2},
\eeq
where $[0,\epsilon)\ni x\mapsto k(x)$ (resp.~$\beta(x)$) is a smooth family of Riemannian metrics (resp.~smooth functions) on $\p\Sigma$.

\begin{definition} One says that $(X,g)$ is even (in the sense of Guillarmou) modulo $\cO(x^3)$ if near $\p X$,
\beq\label{eq:hx}
h(x)=h_0 + x^2 h_1 + \cO(x^3)
\eeq
for some metric $h_0$ and some two-tensor $h_1$ on  $\p X$. 
\end{definition} 

Note that in the standard static setting, if \eqref{eq:hx} holds true then also the Riemannian manifold $(\Sigma,k)$ is even modulo $\cO(x^3)$, i.e. near $\p\Sigma$ we have
\[
k(x)=k_0 + x^2 k_1 + \cO(x^3)
\]
for some metric $k_0$ and two-tensor $k_1$ on  $\p\Sigma$.

\subsection{Klein-Gordon equation in the static model}\label{ss:staticmodel} Suppose that $(X,g)$ is standard static and even modulo $\cO(x^3)$. Then near $\p X$, the Klein-Gordon operator is of the form
\[
P = (-x \p_x + n-1+x e(x)) x\p_x + x^2 \Box_h + \mass,  
\]
where $x\mapsto e(x)$ is a smooth family of functions such that
\[
e(x)=x e_0 + \cO(x^2)
\]
for some $e_0\in\cf(\p \Sigma)$. Following \cite{gannot} (with the addition of $\beta^{\12}$ factors) we consider the operator
\[
\widetilde{P}\defeq \beta^{\frac{1}{2}}x^{-\frac{n}{2}-1} P x^{\frac{n}{2}-1}\beta^{\frac{1}{2}}.
\]
Recall that $P$ is formally self-adjoint with respect to $L^2(X)=L^2(X, g)$, and so $x^{-\frac{n}{2}-1} P x^{\frac{n}{2}-1}$ is formally self-adjoint with respect to $L^2(X,\tilde g)=x^{-\frac{n}{2}}L^2(X)$. The $\beta^{\12}$ factors are useful to eliminate the coefficient in front of $\p_t^2$. One gets indeed that near $\pX$,
\[
\bea
\widetilde{P}&=\beta^{\12}(-\p_x^2 + (\nu^2-\textstyle\frac{1}{4})x^{-2}+ (x e_0 + \cO(x^2)) \p_x + \Box_h )\beta^{\12}\\
&= \p_t ^2 +\beta^{\12}\big(-\p_x^2 + (\nu^2-\textstyle\frac{1}{4})x^{-2}+ (x e_0 + \cO(x^2)) \p_x - \Delta_{k(x)}\big)\beta^{\12}. 
\eea
\]
By setting
\beq\label{eq:defB}
\widetilde{P}\eqdef \p_t^2 + A,
\eeq
or more correctly, $Av\defeq \e^{-\i t \lambda} (\widetilde{P} \e^{\i t\lambda}v)\traa{\{\lambda=0\}}$ for $v\in\cf_\c(\Sigma^\inti)$,   we obtain an  operator $A$ acting on $\cf_\c(\Sigma^\inti)$. It is a positive operator in the sense of the inner product of
\[
L^2(\Sigma)\defeq L^2(\Sigma,dx^2+k).
\] 
For simplicity, in what follows we assume that there exists $C>0$ s.t. $C\leq \beta\leq C^{-1}$ and that $A>0$, i.e. $A$ is \emph{strictly positive}.

Let us consider  the Friedrichs extension of $A$ (in the sense of the Hilbert space $L^2(\Sigma)$), and denote it by the same letter. 

We remark that for $\nu\geq 1$ one expects $A$ to be essentially self-adjoint on $\cf_\c(\Sigma^\inti)$, whereas for $0< \nu<1$ several self-adjoint extensions exist, and in both cases the Friedrichs extension accounts for Dirichlet boundary conditions (see the analysis in \cite{gannot}, cf.~\cite{ishibashiwald,bachelot4,DF} for the case of exact $\AdS$, and also \cite{wald1} for a general argument that explains how the Friedrichs extension corresponds to Dirichlet boundary conditions). The essential feature of the operator $A$ is the presence of the term $-\p_x^2 + (\nu^2-\textstyle\frac{1}{4})x^{-2}$, which has the consequence that many properties of $A$ can be traced back to those of the Schr\"odinger operator with an inverse-square potential considered on the half-line (though here only the behaviour close to $x=0$ is relevant), see e.g.~\cite{truc,DR} for recent results. 
 
The form domain of $A$ (which equals $\Dom A^\12$) is by construction the completion of $\cf_\c(\Sigma^\inti)$ with respect to the norm
\beq\label{eq:lwkjl}
\| v \|^2_{A^{\12}}\defeq  (v | A v)_{L^2(\Sigma)}+  ( v| v)_{L^2(\Sigma)}.
\eeq
Gannot studied in  \cite{gannot} spaces with norms that are equivalent to \eqref{eq:lwkjl}. In particular it follows from \cite[Lem.~3.3]{gannot} (and the subsequent discussion on general manifolds) that any $u$ supported close to the boundary belongs to $\Dom A^\12$ iff it belongs to the `supported' Sobolev space $\dot H^1(\Sigma)$ (defined as the closure of $\cf_\c(\Sigma^\inti)$ with respect to the $H^1$ norm on $\Sigma$).

We consider the \emph{energy space} $\cH_{\rm en}=\Dom A^\12 \oplus L^2(\Sigma)$ with its norm
\[
\| (u_0,u_1) \|_{\cH_{\rm en}}  = \| A^{\12} u_0\|^2_{L^2(\Sigma)} + \|u_1\|^2_{L^2(\Sigma)}.
\]
In this Hilbert space, the operator
\[
H=\begin{pmatrix} 0 & \one \\ A & 0 \end{pmatrix}, \ \ \Dom H = \Dom A \oplus \Dom A^{\12}
\]
is self-adjoint. Using the relation between the two equations $(\i\p_t+H)\phi(t)=0$ and $(\p_t^2+A)u(t)=0$ (namely, $\phi=(u,\i^{-1}\p_t u)$), one concludes in the standard way the well-posedness of the Cauchy problem
\beq\label{eq:stcauchy}
\begin{cases}
\widetilde P u =0,\\
(u,\i^{-1}\p_t u)\traa{t=t_0}=(u_0,u_1)  
\end{cases}
\eeq
in $u\in \cC^0(\rr_t;\Dom A^\12)\cap \cC^1(\rr_t;L^2(\Sigma))$ for any $t_0\in\rr$ and $u_0\in\Dom A^\12$, $u_1\in L^2(\Sigma)$.

\begin{proposition}\label{propsdlksdf}  Suppose $(X,g)$ is standard static and even modulo $\cO(x^3)$. Suppose there exists $C>0$ s.t. $C\leq \beta\leq C^{-1}$ and that $A>0$. Then the Dirichlet retarded/advanced propagator of $P$ is 
\[
P^{-1}_\pm= x^{\frac{n}{2}-1}\beta^{-\frac{1}{2}}\widetilde{P}^{-1}_{\pm}\beta^{-\frac{1}{2}}x^{-\frac{n}{2}-1},
\]
where 
\beq\label{eq:sdflksfl}
(\widetilde{P}^{-1}_\pm f)(t) = \pm\int_{\rr} \theta(\pm (t-s)) \frac{\sin ((t-s)A^{\12} )}{A^{\12} } f(s) ds,
\eeq
$\theta$ being the Heaviside step function.
\end{proposition}
\proof Let us denote
\[
\widetilde{P}^{-1}_{\pm,{\rm Va}}= x^{-\frac{n}{2}+1}\beta^{\frac{1}{2}} P^{-1}_\pm \beta^{\frac{1}{2}}x^{\frac{n}{2}+1}.
\]
We focus on the  `$+$' case, the `$-$' case being analogous.  We want to show that 
\beq\label{telwerw}
\widetilde{P}^{-1}_{+,{\rm Va}} f -\widetilde{P}^{-1}_{+} f=0
\eeq
for all $f$ belonging to some dense subspace of $x^{-\frac{n}{2}-1}\Hl^{-1,\infty}(X)$, for instance $f\in \dot\cC_\c^\infty(X)$. Since we have
\[
\widetilde P  (\widetilde{P}^{-1}_{+,{\rm Va}} f - \widetilde{P}^{-1}_{+} f ) =0,
\]
and $\widetilde{P}^{-1}_{+,{\rm Va}} f - \widetilde{P}^{-1}_{+} f$ has vanishing Cauchy data in the past of $\supp f$, we can conclude  \eqref{telwerw} from the uniqueness of the solution of the Cauchy problem \eqref{eq:stcauchy} provided that we first check that
\beq\label{eq:lfkjeflk}
\widetilde{P}^{-1}_{+,{\rm Va}} f,\widetilde{P}^{-1}_{+} f\in \cC^0(\rr_t;\Dom A^\12)\cap \cC^1(\rr_t;L^2(\Sigma)).
\eeq

Let us show the first assertion. By the mapping properties of the Dirichlet retarded propagator $P_+^{-1}$ and the uniform boundedness of $\beta$ and $\beta^{-1}$, 
\[
\widetilde{P}^{-1}_{+,{\rm Va}}f\in x^{-\frac{n}{2}+1} \Hl^{1,\infty}(X).
\]
From the definition of $\Hb^{1,\infty}(X)$ and the relation 
\beq\label{eq:rel1}
x^{-n/2} L^2(X)=L^2(X,\tilde g)=L^2(\rr_t;L^2(\Sigma))
\eeq
we obtain
\beq\label{eq:rel2}
x^{-\frac{n}{2}+1} \Hl^{1,\infty}(X)\subset x \cf(\rr_t;L^2(\Sigma))\cap \cf(\rr_t;\dot H^1(\Sigma)).
\eeq
In view of the result mentioned below \eqref{eq:lwkjl}, i.e.~the equivalence of $\dot H^1(\Sigma)$ and $\Dom A^{\12}$ close to the boundary, this yields the first part of \eqref{eq:lfkjeflk}.

The proof of the second assertion in \eqref{eq:lfkjeflk} is straightforward using \eqref{eq:sdflksfl}. 
\qeds

In the setup of Proposition \ref{propsdlksdf}, the construction of quantum fields (corresponding to the ground state for the static dynamics) is standard, see e.g.~\cite[Sec.~18.3]{derger}. For later reference we give below a lemma on two-point functions (this terminology is explained in the next section). 

\begin{lemma}\label{lemma:sttwom} Let $(X,g)$ be as in Proposition \ref{propsdlksdf} and  suppose $A\geq m^2\one$ for some $m>0$. Let
\[
\Lambda^\pm= x^{\frac{n}{2}-1}\beta^{-\frac{1}{2}}\widetilde{\Lambda}^{\pm}\beta^{-\frac{1}{2}}x^{-\frac{n}{2}-1},
\]
where 
\beq\label{eq:sdflksfl2}
(\widetilde{\Lambda}^\pm f)(t) = \int_{\rr} A^{-\12} \e^{\pm\i(t-s)A^{\12}} f(s) ds
\eeq
for $f\in\cfd_\c(X)$. Then $\Lambda^\pm$ extends to a continuous map $\Lambda^\pm:\Hc^{-1,\infty}(X)\to \Hl^{1,\infty}(X)$ such that $\Lambda^+-\Lambda^-=\i (P_+^{-1}-P_-^{-1})$ and $( f | \Lambda^\pm f)_{L^2}\geq 0$ for all $f\in  \Hc^{-1,\infty}(X)$. Furthermore, $\Lambda^\pm:x^{\frac{n}{2}+1} L^1(\rr_t;L^2(\Sigma))\to x^{\frac{n}{2}-1} \cC^1_{\rm bd}(\rr_t;L^2(\Sigma))$, where the $\rm bd$ subscript refers to boundedness in $t$. Denoting $D_t=\i^{-1}\p_t$,
\beq\label{eq:sflksjf}
\chi_\mp(D_t)\Lambda^\pm =0
\eeq
on $x^{\frac{n}{2}+1} L^1(\rr_t;L^2(\Sigma))$ for all $\chi_\pm\in \cf(\rr)$ such that $\chi_\pm=0$ in a neighborhood of $\pm(-\infty,m^2]$ and $\chi_\pm=1$ on $\pm[m^2+1,\infty)$.
\end{lemma}
\proof Using the definition of $\widetilde{\Lambda}^\pm$ and $\widetilde{P}_{\pm}^{-1}$ we can write 
\[
{\widetilde{\Lambda}}^\pm=(\p_t\otimes A^{-\12})( \widetilde{P}_+^{-1}- \widetilde{P}_-^{-1})\pm\i( \widetilde{P}_+^{-1}- \widetilde{P}_-^{-1}),
\]
as operators on $L^1(\rr_t;L^2(\Sigma))$. Correspondingly, from the definition of $\Lambda^\pm$ and Proposition \ref{propsdlksdf} we deduce
\beq\label{eq:bububub}
{\Lambda}^\pm=(\p_t\otimes x^{\frac{n}{2}-1}A^{-\12}x^{-\frac{n}{2}+1})( P_+^{-1}- P_-^{-1})\pm\i( P_+^{-1}- P_-^{-1})
\eeq
on $x^{\frac{n}{2}+1}L^1(\rr_t;L^2(\Sigma))$. To show that \eqref{eq:bububub} defines an operator that maps continuously $\Hc^{-1,\infty}(X)\to \Hl^{1,\infty}(X)$, in view of the mapping properties of $P_\pm^{-1}$ and $\p_t\otimes\one$ it suffices to prove that 
\beq\label{eq:ddskld}
(\one\otimes A^{-\12}):x^{-\frac{n}{2}+1}\Hl^{1,\infty}(X)\to x^{-\frac{n}{2}+1}\Hl^{1,\infty}(X)
\eeq
continuously. We first note that $A^{-\12}$ maps $\Dom A^{\12}$ to $\Dom A$ and similarly, 
\beq\label{eq:direct1}
(\one\otimes A^{-\12}): \cf(\rr_t;\Dom A^{\12})\to \cf(\rr_t;\Dom A).
\eeq
By \cite[Lem.~3.3 \& Sect.~3.10]{gannot}, $\Dom A  \subset xL^2(\Sigma)\cap \dot H^1(\Sigma)$. Using this and the relations between the various spaces stated in \eqref{eq:rel1}, we conclude 
\beq\label{eq:ddskld2}
(\one\otimes A^{-\12}):x^{-\frac{n}{2}+1}H_{0,\loc}^1(X)\to x^{-\frac{n}{2}+1}H_{0,\loc}^1(X).
\eeq
Furthermore, by similar arguments, $(\one\otimes A^{-1})$ restricts to a positive-definite bounded operator on $x^{-\frac{n}{2}+1}H_{0,\loc}^1(X)$, the square root of which is \eqref{eq:ddskld2}. Thus, \eqref{eq:ddskld} can be concluded from 
the boundedness statement
\[
(\one\otimes A^{-1}):x^{-\frac{n}{2}+1}\Hl^{1,\infty}(X)\to x^{-\frac{n}{2}+1}\Hl^{1,\infty}(X).
\]
The latter follows from \cite[Thm.~3]{gannot} and the remark on different spaces of conormal distributions preceding \cite[Lem.~4.15]{gannot}.

The remaining assertions are proved by direct computations using \eqref{eq:sdflksfl2}. 
\qeds

Finally, we will need an auxiliary lemma which states that an asymptotically $\AdS$ spacetime can be deformed to one that contains a standard static region.

\begin{lemma}\label{lem:deform} Suppose $(X,g)$ is an asympotically $\AdS$ spacetime and assume $({\rm TF})$ and $({\rm PT})$. For any $t_2\in\rr$ there exists a static asymptotically $\AdS$ spacetime  $(X,g')$ and $t_0<t_1<t_2$ such that $g'$ equals $g$ on $\{t\geq t_1\}$ and the region $\{t\leq t_0\}$ of $(X,g')$ has an extension to some standard static asymptotically $\AdS$ spacetime which is even modulo $\cO(x^3)$ and in which $C\leq \beta \leq C^{-1}$ for some $C>0$.
\end{lemma}
\proof Since $X=\rr\times\Sigma$, we can construct $g_{\rm st}$ such that $(X,g_{\rm st})$ is standard static and even modulo $\cO(x^3)$, with the boundary-defining function of $\p X$ being defined using the boundary-defining function of $\p\Sigma$. We denote this boundary-defining function by $x_{\rm st}$. Next, we define $g'=g_{\rm st}$ on $\{t\leq t_0\}$ and $g'=g$ on $\{t\geq t_1\}$. Similarly, we set  $g'=g_{\rm st}$ on $\{t\leq t_0\}$ and $g'=g$ on $\{t\geq t_1\}$. The definition of $x'$ can be extended to the intermediate region $\{t_0<t<t_1\}$ as to yield a boundary defining function of $\p X$. Then we extend the metric $h'\defeq x'^2 g'$ to the intermediate region. By setting $g'=(x')^{-2}(-(dx')^2+h')$ we obtain an asymptotically $\AdS$ spacetime $(X,g')$.\qed

\section{Singularities of propagators and two-point functions}\label{sec:propagators}

\subsection{Operator $\b$-wave front set}\label{ss:WFb}

Let us fix some $k_1,k_2\in\zz$.

We denote by $\cWb$ the set of bounded operators  from  $\Hc^{k_2,-\infty}(X)$ to $\Hl^{k_1,\infty}(X)$. Elements of $\cWb$ will play the r\^ole of regularizing operators. Note that $\Psi_\b^{-\infty}(X)\subset\cWb$.

For the sake of brevity, if $E,F$ are topological spaces, we write $\Lambda: E\to F$ to mean that $\Lambda$ is continuous. 

Below we introduce an operatorial $\b$-wave front set which is a subset of $(\be T^* X\setminus\!\zero)\times(\be T^* X\setminus\!\zero)$. As such, it gives no information about certain kinds of singularities (potentially located at $\zero\times \be T^* X$ or $\be T^* X\times\zero$), it will however turn out satisfactory for our purposes.

\begin{definition} \label{def:wfbp} Suppose $\Lambda:\Hc^{k_2,-\infty}(X)\to\Hl^{k_1,-\infty}(X)$. We say that $(q_1,q_2)\in (\be T^* X\setminus\!\zero)\times(\be T^* X\setminus\!\zero)$ is not in $\wf'_\b(\Lambda)$ if there exist  $B_i\in \Psi_\b^0(X)$, elliptic at $q_i$ ($i=1,2$), and such that $B_1 \Lambda B_2^*\in \cWb$.
\end{definition}

Since $\wf'_\b(\Lambda)$ is invariant under the componentwise, fiberwise $\rr_+$-action of dilations, we may replace each copy of $\be T^* X\setminus\!\zero$ by the quotient
\[
\be S^*X\defeq (\be T^*X\setminus\zero) / \rr_+
\] 
by the fiberwise $\rr_+$-action of dilations. We will often do so without stating it explicitly; this is especially useful when discussing neighborhoods. \medskip

For $B\in\Psi^s(X)$ there is another natural notion of operator wave front set, denoted here $\wf^\Psi_\b(A)$, which describes where in $\be T^*X \setminus \zero$ the symbol of $B$ is not of order $-\infty$, see Appendix \ref{appb} for the precise definition (we stress that we use non-standard notation, as  $\wf'_\b$ is usually reserved for the pseudo-differential operator $\b$-wave front set which we denote here by $\wf^\Psi_\b$). There is a simple relation between the two operator wave front sets.

\begin{lemma} If $B\in\Psi^s(X)$ then $\wf'_\b(B)= \{ (q,q) : \ q\in \wf^\Psi_\b(B) \}$.
\end{lemma}
\proof Let us recall that
\[
\forall \, A,B\in\Psi_\b(X), \   \wf^\Psi_\b(AB)\subset\wf^\Psi_\b(A)\cap \wf^\Psi_\b(B).
\]
If $q_1,q_2\in T^*X\setminus \zero$ and $q_1\neq q_2$, one can choose $A_i$ elliptic at $q_i$ such that $\wf^\Psi_\b(A_1)\cap \wf^\Psi_\b(A_2^*)=\emptyset$. Then $A_1 B A_2^*\in \Psi^{-\infty}(X)$ and so $(q_1,q_2)\in\wf'_\b(B)$. This proves that $\wf'_\b(B)$ lies on the diagonal in $\be S^*X \times \be S^*X$.

Suppose now $q\notin \wf^\Psi_\b(B)$. Then $A_1 B \in\Psi_\b^{-\infty}(X)$ for some $A_1$ elliptic at $q$, hence $(q,q_2)\notin \wf'_\b(B)$ for any $q_2\in  T^*X\setminus \zero$. This proves `$\subset$'.

On the other hand, suppose that $(q,q)\notin\wf'_\b(B)$, so that there exists $A_1,A_2$ elliptic at $q$ such that $R\defeq A_1 B A_2^*\in\Psi_\b^{-\infty}(X)$. Let $G$ be a parametrix of $A_2^*$. Then there exists a neighbourhood $\Gamma\subset \be S^*X$ of $q$ such that $\wf_\b(\one-A_2^* G)\cap \Gamma=\emptyset$. Furthermore, there exists $A_3$ elliptic at $q$ satisfying $\wf^\Psi_\b(A_3)\subset \Gamma$. This gives    
\[
A_3 A_1 B = A_3 R G  + A_3 A_1 B (\one-A_2^* G) \in \Psi_\b^{-\infty}(X). 
\]
Since $A_3 A_1$ is elliptic at $q$ this gives $q\notin \wf^\Psi_\b(B)$, which proves  `$\supset$'.\qed

\begin{lemma}\label{lem:altdefwfb} For any $q_1,q_2\in \be S^* X$, $(q_1,q_2)\notin\wf'_\b(\Lambda)$ if and only if there exist neighbourhoods $\Gamma_i$ of $q_i$ such that for all $B_i\in\Psi_\b^0(X)$ elliptic at $q_i$ satisfying $\wf^\Psi_\b(B_i)\subset \Gamma_i$, $i=1,2$, $B_1 \Lambda B_2^*\in\cWb$.
\end{lemma}
\proof Suppose $(q_1,q_2)\notin\wf'_\b(\Lambda)$, so that there exists $A_i\in\Psi_\b^0(X)$, $i=1,2$,  elliptic at $q_i$, such that $A_1\Lambda A_2^* \in \cWb$. There exists a compact neighbourhood $\Gamma_i$ of $q_i$ on which $A_i$ is elliptic. Therefore, there exists $A_i^\inv\in\Psi^0_\b(X)$ such that
\[
\wf^\Psi_\b(A^\inv_i A_i - \one)\cap \Gamma_i = \emptyset.
\]
Let $B_i\in\Psi_\b^0(X)$ be elliptic at $q_i$ and such that $\wf^\Psi_\b(B_i)\subset \Gamma_i$. These conditions on the wave front sets imply that
\beq\label{eq:bababa}
B_1(A^\inv_1 A_1 - \one)\in\Psi_\b^{-\infty}(X), \ \ (A_2^* (A^\inv_2)^* - \one)B_2^*\in\Psi_\b^{-\infty}(X).
\eeq
We can write
\[
\bea
B_1 \Lambda B_2^*&=B_1 A_1^\inv A_1 \Lambda  A_2^* (A^\inv_2)^* B_2^* + B_1 (\one - A_1^\inv A_1) \Lambda A_2^* (A^\inv_2)^* B_2^*\\
 & \phantom{=}\, + B_1 A_1^\inv A_1  \Lambda (\one-A_2^* (A^\inv_2)^* )B_2^* \\
 & \phantom{=}\, + B_1 (\one - A_1^\inv A_1) \Lambda (\one-A_2^* (A^\inv_2)^*)B_2^*.
\eea 
\]
By $A_1\Lambda A_2^* \in \cWb$ and \eqref{eq:bababa}, all the summands belong to $\cWb$, hence $B_1 \Lambda B_2^*\in\cWb$. 

The opposite direction is trivial.\qed

\begin{lemma}\label{wfs} Let $\Lambda,\tilde\Lambda:\Hc^{k_2,-\infty}(X)\to\Hl^{k_1,-\infty}(X)$, then
\[
\wf_\b'(\Lambda+\tilde\Lambda)\subset \wf_\b'(\Lambda)\cup \wf_\b'(\tilde\Lambda).
\]
\end{lemma}
\proof If $(q_1,q_2)\notin \wf_\b'(\Lambda)$ and $(q_1,q_2)\notin \wf_\b'(\tilde\Lambda)$ then by Lemma \ref{lem:altdefwfb} we can choose $B_1,B_2$ elliptic at resp. $q_1,q_2$ such that both $B_1 \Lambda B_2^*$ and $B_1 \tilde\Lambda B_2^*$ belong to $\cWb$. Hence  $B_1 (\Lambda+\tilde\Lambda) B_2^*$ belongs to $\cWb$ and thus $(q_1,q_2)\notin \wf_\b'(\Lambda+\tilde\Lambda)$.\qed

\begin{proposition} Suppose $\wf'_\b(\Lambda)=\emptyset$. Then $\Lambda\in\cWb$.
\end{proposition}
\proof The proof is an adaptation of \cite[Lem.~3.10]{corners} to the case of the operator wave front set. It suffices to show that for any $p_1,p_2\in X$ there exists $\phi_1,\phi_2\in\cf_\c(X)$ with $\phi_i\equiv 1$ near $p_i$ such that $\phi_1 \Lambda \phi_2\in\cWb$.

By definition of $\wf'_\b(\Lambda)$, for any $q,q'\in \be S^* X$ there exist $B_{1,q},B_{2,q'}\in\Psi_\b^0(X)$ elliptic at resp. $q$, $q'$, such that $B_{1,q} \Lambda B_{2,q'}^*\in\cWb$. Let $\Gamma_{1,q}$ be the set on which $B_{1,q}$ is elliptic. 

 Then $\{ \Gamma_{1,q} : \ q\in \be S^*_{p_1} X \}$ is an open cover of $ \be S^*_{p_1} X$. By compactness we can find a finite subcover $\{ \Gamma_{1,q_j}\}_{j=1}^{N}$. Then $B_1=\sum_{j} B_{1,q_j}^*B_{1,q_j}\in\Psi^0_\b(X)$ is elliptic on $\be S^*_{p_1} X$ (this follows from $\sigma_{\b,0}(B_1)$ being equal to $\sum_{j} |\sigma_{\b,0}(B_{1,q_j})|^2$). In a similar way we construct $B_2= \sum_{l} B_{2,q'_l}^* B_{2,q'_l}\in \Psi_\b^0(X)$ elliptic on  $\be S^*_{p_2} X$. This gives
\[
B_1 \Lambda B_2 = \textstyle\sum_{j,l}  B_{1,q_j}^*B_{1,q_j} \Lambda B_{2,q'_l}^* B_{2,q'_l}\in \cWb
\]
using that the sum is finite. 

We can find a microlocal parametrix of $B_1$ and $B_2$, i.e.~$B_i^\inv\in\Psi_\b^0(X)$ such that $R_1=\one-B_1^{\inv} B_1 $ and $R_2=\one-B_2 B_2^{\inv}$ satisfy $\wf_\b'(R_i)\cap \be S^*_{p_i} X=\emptyset$. This implies that there is a neighborhood $O_i$ of $p_i$ in $X$ such that $\wf'_\b(R_i)\cap \be S^*_{O_i} X=\emptyset$. Let $\phi_i\in\cf_\c(X)$ be such that $\supp \phi_i\subset O_i$ and $\phi_i\equiv 1$ near $p_i$. We have
\[
\bea
 \phi_1 \Lambda\phi_2 &=\phi_1 B_1^\inv (B_1 \Lambda  B_2) B^\inv_2 \phi_2+ \phi_1 R_1 \Lambda B_2 B^\inv_2 \phi_2\\
 & \phantom{=}\, + \phi_1 B_1^\inv B_1  \Lambda R_2 \phi_2 +  \phi_1 R_1 \Lambda R_2 \phi_2,
\eea 
\]
where all the summands belong to $\cWb$, hence $\phi_1 \Lambda\phi_2\in \cWb$. \qeds

Thus, $\wf'_\b(\Lambda)=\emptyset$ implies in particular that for any $\psi_1,\psi_2\in\cf(X)$ supported away from the boundary $\p X$, the Schwartz kernel of $\psi_1 \Lambda \psi_2$ is smooth as a distribution on $X^\inti\times X^\inti$.

In the next lemma we take $k_1=-k_2\eqdef k$.

\begin{lemma}\label{lempos} Suppose that $\Lambda:\Hc^{-k,-\infty}(X)\to\Hl^{k,-\infty}(X)$ and $\Lambda\geq 0$. If $(q_1,q_2)\in \wf'_\b(\Lambda)$ for some $q_1,q_2\in T^*X\setminus\zero$ then $(q_1,q_1)\in \wf'_\b(\Lambda)$ or $(q_2,q_2)\in \wf'_\b(\Lambda)$.
\end{lemma}
\proof Suppose $(q_1,q_1),(q_2,q_2)\notin \wf'_\b(\Lambda)$. By Lemma \ref{lem:altdefwfb} we can find $B_i$ elliptic at $q_i$ such that $B_i \Lambda B_i^*: \Hc^{-k,-\infty}(X)\to\Hl^{k,\infty}(X)$. Since $\Hc^{-k,-\infty}(X)$ is dual to $\Hl^{k,\infty}(X)$, this implies that 
\[
\sup_{f\in\cU,f_i\in\cU_i}\left|( f| B_i\Lambda B_i^* f_i )_{L^2}\right| < \infty,
\]
for all bounded subsets $\cU$, $\cU_i$ of $\Hc^{-k,-\infty}(X)$. Using the Cauchy-Schwarz inequality for the positive sesquilinear form associated with $\Lambda$, we obtain
\[
\sup_{f_i\in\cU_i}| ( f_1| B_1 \Lambda B_2^* f_2)_{L^2}| \leq \sup_{f_1\in\cU_1}( f_1| B_1 \Lambda B_1^* f_1)^\12_{L^2} \sup_{f_2\in\cU_2}( f_2| B_2 \Lambda B_2^* f_2)^\12_{L^2}< \infty.
\]
This implies that $B_1 \Lambda B_2^*$ maps continuously $\Hc^{-k,-\infty}(X)$ to $\Hl^{k,\infty}(X)$, and thus that $(q_1,q_2)\notin \wf'_\b(\Lambda)$. 
\qed

\subsection{Propagators and two-point functions} We now introduce the concepts relevant for non-interacting QFT (here only scalar fields are considered) on an asymptotically $\AdS$ spacetime $(X,g)$. We recall that $G=P_+^{-1}- P_{-}^{-1}$ is the Dirichlet causal propagator.

\begin{definition} We say that  $\Lambda^\pm : \Hc^{-1,-\infty}(X)\to \Hl^{1,-\infty}(X)$ are \emph{two-point functions} if
\beq\label{eq:2ptfct}
\bea i) \quad & P\Lambda^\pm = \Lambda^\pm P =0,\\
ii) \quad & \Lambda^+-\Lambda^- = \i G \,\mbox{ and }\, \Lambda^\pm\geq 0.
\eea
\eeq
\end{definition}

By duality, $\Lambda^\pm : \Hc^{-1,\infty}(X)\to \Hl^{1,\infty}(X)$. These conditions ensure thus that $\Lambda^\pm$ induce well-defined positive sesquilinear forms on the symplectic space $\Hc^{-1,\infty}(X)/P\Hc^{1,\infty}(X)$.  Once $\Lambda^\pm$ are given, the standard apparatus of algebraic QFT can be used to construct quantum fields, see e.g.~\cite{derger,KM}. We emphasize that we use the non-standard conventions borrowed from the complex formalism, see \cite{GOW} for the relation between two-point functions $\Lambda^\pm$, states and fields. 

Just as on globally hyperbolic spacetimes, one does not expect all two-point functions to be physical. In the present setup we propose the following definition, which essentially reduces to the well-established Hadamard condition in the bulk, but which also involves conormal regularity at the boundary.

We use Definition \ref{def:wfbp} with $k_1=1$, $k_2=-1$ for the primed $\b$-wave front set.

\begin{definition} We say that $\Lambda^\pm : \Hc^{-1,-\infty}(X)\to \Hl^{1,-\infty}(X)$  are \emph{holographic Hadamard two-point functions} if they satisfy \eqref{eq:2ptfct} and
\beq\label{eq:holhad}
\wf_\b'(\Lambda^\pm )\subset \dot\cN^\pm\times \dot\cN^\pm.
\eeq
\end{definition}

The property \eqref{eq:holhad} will be called the \emph{holographic Hadamard condition} in view of the conormal regularity it implies.

If $q_1,q_2\in \be S^*X$, we write $q_1\dot\sim q_2$ if $q_1,q_2\in \dot \cN$ and $q_1,q_2$ can be connected by a generalized broken bi-characteristic. 

We will need an operatorial version of Vasy's propagation of singularities theorem.

\begin{proposition}\label{opprop} Let $\Lambda:\Hc^{-1,-\infty}(X)\to \Hl^{1,-\infty}(X)$ and suppose $(q_1,q_2)\in\wf'_\b(\Lambda)$. If $P\Lambda=0$, then $q_1\in \dot\cN$, and $(q_1',q_2)\in\wf'_\b(\Lambda)$ for all $q_1'$ such that $q_1'\dot\sim q_1$. Similarly, if $\Lambda P=0$ then $q_2\in \dot\cN$, and $(q_1,q_2')\in\wf'_\b(\Lambda)$   for all $q_2'$ such that $q_2'\dot\sim q_2$.
\end{proposition}
\proof For the first statement, suppose $(q_1,q_2)\notin \wf'_\b(\Lambda)$. Then by definition there exist $B_1,B_2\in\Psi_\b^0(X)$ elliptic at respectively $q_1,q_2$ such that for any bounded subset $\cU\subset\Hc^{-1,-\infty}(X)$, the set $B_1 \Lambda B_2^* \cU$ is bounded in $\Hl^{1,\infty}(X)$. By propagation of singularities applied to $\Lambda B_2^* \cU$, using the fact that the estimates underpinning Vasy's theorem are \emph{uniform}, one deduces the existence of $B_1'\in\Psi_\b^0(X)$ elliptic at $q_1'$ such that $B_1' \Lambda B_2^* \cU$ is bounded in $\Hl^{1,\infty}(X)$, hence $(q_1',q_2)\notin\wf'_\b(\Lambda)$.

To see that the second statement is true, let us observe that if $B_1 \Lambda B_2^*$ is regularizing then so is $B_2 \Lambda^* B_1^*$, and furthermore, $\Lambda^* B_1^*$ satisfies $ P \Lambda^* B_1^*=0$. This way the proof can be reduced to the previous case. 
\qeds

If we fix some $t_1,t_2\in\rr$, $t_1<t_2$, by assumption $({\rm TF})$ all $\GBB$s reach the region of $X$ where $t\in [t_1,t_2]$. Thus, we obtain as an immediate corollary of Proposition \ref{opprop} (note that by definition of $\wf'_\b$, the statement below says nothing about potential singularities located at $\zero\times \be T^*X$ or $\be T^*X\times\zero$):

\begin{lemma}\label{slkmdff} Suppose that $\Lambda^\pm$ is a pair of two-point functions that satisfy \eqref{eq:holhad} in the region $\{ t_1 \leq t\leq t_2 \}$. Then $\Lambda^\pm$ are holographic Hadamard two-point functions, i.e.~they satisfy \eqref{eq:holhad} everywhere.
\end{lemma}

This allows us to prove the existence of holographic Hadamard two-point functions in analogy to the deformation argument of Fulling, Narcowich and Wald, formulated originally in the setting of globally hyperbolic spacetimes \cite{FNW}.

\begin{theorem}\label{thm:existence} Suppose $(X,g)$ is an asymptotically $\AdS$ spacetime and assume $({\rm TF})$, $({\rm PT})$ and $\nu>0$. Then there exist   holographic Hadamard two-point functions on $(X,g)$.
\end{theorem}
\proof We first claim that it suffices to construct a pair of operators $\Lambda^\pm$ acting on $H^{-1,-\infty}_{0,\b,[t_1,t_2]}(X)$ (recall that this is the space of all $\Hc^{-1,-\infty}(X)$ supported in $\{t_1\leq  t\leq t_2\}$), such that $\Lambda^\pm$ satisfy all the conditions required of holographic Hadamard two-point functions with $\Hc^{-1,-\infty}(X)$ replaced by $H^{-1,-\infty}_{0,\b,[t_1,t_2]}(X)$ (and with an estimate on $\wf'_\b(\Lambda^\pm)$ only above $\{t_1\leq t\leq t_2\}$). Indeed, we can always continuously extend such $\Lambda^\pm$ to $\Hc^{-1,-\infty}(X)$ using Proposition \ref{tslice} (the so-called `time-slice property'). Namely, the extension is defined by
\beq\label{eq:asdkjln}
(\imath_{t_1,t_2}^{-1})^*\Lambda^\pm\imath_{t_1,t_2}^{-1}: \Hc^{-1,\infty}(X)\to \Hl^{1,\infty}(X),
\eeq
which then extends to $\Hc^{-1,-\infty}(X)$ by duality. We can check that this is indeed a pair of two-point functions: positivity is obvious, furthermore,
\[
(\imath_{t_1,t_2}^{-1})^*\Lambda^+\imath_{t_1,t_2}^{-1}-(\imath_{t_1,t_2}^{-1})^*\Lambda^-\imath_{t_1,t_2}^{-1}=\i (\imath_{t_1,t_2}^{-1})^*G\imath_{t_1,t_2}^{-1}=\i G
\]
using Proposition \ref{tslice} (or, equivalently, using the formula for $\imath_{t_1,t_2}^{-1}$).  
The holographic Hadamard condition is then satisfied by \eqref{eq:asdkjln}  in view of Lemma \ref{slkmdff}.

Since by the above argument, the problem of proving existence is reduced to an arbitrary compact time interval, we can assume without loss of generality that the spacetime $(X,g)$ has a standard static region $\{t \leq t_0 \}$, $t_0<t_1$, as in Lemma \ref{lem:deform}.
We observe that Vasy's propagation of singularities result is unaffected if one adds to $P$ a smooth potential $V>0$ that depends only on $t$, thus we can also assume without loss of generality that $A\geq m^2>0$ in  $\{t \leq t_0 \}$ (recall that the operator $A$ was defined Subsection \ref{ss:staticmodel}).
 
 Again, since it suffices to prove the existence in an arbitrary compact time interval, we are reduced to doing so in a standard static region in which $A\geq m^2>0$.

We recall that in the standard static setting we have already constructed two-point functions, subsequently denoted by $\Lambda^\pm_{\rm vac}$, such that (see Lemma \ref{lemma:sttwom})
\beq\label{eq:sflksjf2}
\chi_\mp(D_t)\Lambda^\pm_{\rm vac} =0
\eeq
on $x^{\frac{n}{2}+1} L^1(\rr_t;L^2(\Sigma))$ for all $\chi_\pm\in \cf(\rr)$ such that $\chi_\pm=0$ in a neighborhood of $\pm(-\infty,m^2]$ and $\chi_\pm=1$ on $\pm[m^2+1,\infty)$. By the elliptic regularity statement of Proposition \ref{opprop} and Lemma \ref{lempos},  
\[
\WF'_\b(\Lambda^\pm_{\rm vac})\subset \dot\cN\times\dot\cN.
\]
Next, let $q_1\in\dot\cN^\mp$. Let us denote by $\tau$ the covariable respective to $t$. We can write
\[
\chi_\mp(D_t)\otimes \one = B_1 + B_2 + R_{-\infty},
\]
where $B_1\in \Psi_\b^0(X)$ has a symbol which coincides with $\chi(\tau)$ outside of a small neighborhood $\Gamma\subset\be S^* X$ of $\{\tau = 0\}$ (chosen such that $q_1\notin\Gamma$), $B_2$ is the quantization of a `symbol' supported near $\{\tau = 0\}$, and $R_{-\infty}\in \cW_\b^{-\infty}(X)$. Furthermore, we can find $B\in \Psi_\b^0(X)$ elliptic at $q_1$ such that $B B_2\in \cW_\b^{-\infty}(X)$. From \eqref{eq:sflksjf2} one finds
\beq\label{eq:rgergm}
B B_1 \Lambda^\pm = - (B B_2+ B R_{-\infty}) \Lambda^\pm
\eeq
on a dense subset of $\Hl^{-1,\infty}(X)$, and hence on $\Hl^{-1,\infty}(X)$. The right hand side of \eqref{eq:rgergm} belongs to $\cW_\b^{-\infty}(X)$ and $B B_1$ is elliptic at $q_1$, therefore $(q_1,q_2)\notin \wf'(\Lambda^\pm)$ for any $q_2\in \be S^* X$. Since $q_1\in\dot\cN^\mp$ was arbitrary, using  Lemma \ref{lempos} we can conclude 
\[
\WF'_\b(\Lambda^\pm_{\rm vac})\subset \dot\cN^\pm \times\dot\cN^\pm
\]
as desired.
\qeds

Using the propagation of singularities we can now estimate more precisely the $\b$-wave front set of $\Lambda^\pm$ and of various propagators for $P$.  Let us recall the notation $\pi:\be T^*X \to X$ for the bundle projection.

\begin{theorem}\label{prop:wfs} Suppose $(X,g)$ is as in Theorem \ref{thm:existence} and $\nu>0$. Then:
\beq\label{eq:wfPpm}
\wf'_\b(P_\pm^{-1})\setminus\tdiag\subset \{ (q_1,q_2) : \ q_1 \dot\sim q_2, \ \pm t(\pi q_1)>\pm t(\pi q_2) \},
\eeq
where $\tdiag = \{ (q_1,q_2)\in \be S^* X\times \be S^*X : \  t(\pi q_1)= t(\pi q_2)  \}$. Furthermore, suppose that $\Lambda^\pm$ are holographic Hadamard two-point functions. Then 
\beq\label{eq:mlkfn}
\wf'_\b(\Lambda^\pm)\subset \{ (q_1,q_2)\in\dot\cN^\pm\times\dot\cN^\pm : \ q_1 \dot\sim q_2 \mbox{ or }  \pi q_1= \pi q_2\}. 
\eeq
Moreover, setting $P^{-1}_{\F}\defeq \i^{-1} \Lambda^+ + P_-^{-1}$ and $P^{-1}_{\aF}\defeq -\i^{-1} \Lambda^- + P_-^{-1}$, we have
\beq \label{eq:wfF} 
\bea
\wf'_\b(P_\F^{-1})\setminus \tdiag &\subset \{ (q_1,q_2): \ q_1 \dot\sim q_2, \mbox{ and } \pm t(\pi q_1)\leq \pm t(\pi q_2) \mbox{ if } q_1\in\dot\cN^\pm \},\\
\wf'_\b(P_{\rm \bar{F}}^{-1})\setminus\tdiag&\subset \{ (q_1,q_2): \ q_1 \dot\sim q_2, \mbox{ and } \mp t(\pi q_1)\leq \mp t(\pi q_2) \mbox{ if } q_1\in\dot\cN^\pm \}.
\eea
\eeq
\end{theorem}
\proof From the definition of $P_\pm^{-1}$ it follows that for any $(q_1,q_2)$, if $\pm t(\pi q_1)< \pm t(\pi q_2)$ then we can find $\chi_1,\chi_2\in \cf_\c(X)$ with disjoint supports such that $\chi_i(\pi q_i)\neq 0$, $i=1,2$, and $\chi_1\circ P_\pm ^{-1}\circ\chi_2=0$. Thus, 
\beq\label{eq:newtemp}
 \pm t(\pi q_1)>\pm t(\pi q_2) \ \Longrightarrow \ (q_1,q_2)\notin \wf'_\b(P_\pm^{-1}). 
\eeq
On the other hand, for any  $(q_1,q_2)\in \wf'_\b(P_\pm^{-1})$ such that $\pi q_1 \neq \pi q_2$, by elliptic regularity (more precisely, by Proposition \ref{opprop} applied near $\pi q_1$ to $P_\pm^{-1}\circ \chi_2$, where $\chi_2\in \cf_\c(X)$ is supported in a sufficiently small neighborhood of $q_2$ and $\chi_2(\pi q_2)\neq 0$) we get $q_1\in \dot\cN$, and similarly $q_2\in \dot\cN$.    

We will now show the more precise estimate
\beq\label{eq:moreprecise}
\wf'_\b(P_\pm^{-1})\subset \{ (q_1,q_2): \ q_1 \dot\sim q_2 \mbox{ or }  \pi q_1= \pi q_2  \}. 
\eeq
Suppose $(q_1,q_2)\in\dot\cN\times\dot\cN$ does not satisfy $q_1 \dot\sim q_2$ nor $\pi q_1= \pi q_2$. Then we can find $q_1'$ such that $q_1'\dot\sim q_1$ and $\pm t(\pi q_1')>\pm t(\pi q_2)$. By virtue of \eqref{eq:newtemp}, $(q_1',q_2)\notin\wf'_\b(P_\pm^{-1})$. By propagation of singularities (more precisely, by Proposition \ref{opprop} applied to $P_\pm^{-1}\circ \chi_2$, with $\chi_2\in \cf_\c(X)$ supported in a sufficiently small neighborhood of $q_2$ and such that $\chi_2(\pi q_2)\neq 0$), $(q_1,q_2)\notin\wf'_\b(P_\pm^{-1})$.

We now turn our attention to $\Lambda^\pm$. Since $\Lambda^+ - \Lambda^-=\i (P_+^{-1}-P_-^{-1})$ and $\wf'_\b(\Lambda^+)\cap \wf'_\b(\Lambda^-)=\emptyset$, we have 
\[
\wf'_\b(\Lambda^\pm)\subset (\dot\cN^\pm\times\dot\cN^\pm)\cap \big(\wf'_\b(P_+^{-1})\cup\wf'_\b(P_-^{-1})\big). 
\]
In view of \eqref{eq:moreprecise} this yields \eqref{eq:mlkfn}.

Let us now estimate the wave front set of $P_{\rm F}^{-1}= \i^{-1} \Lambda^+ + P_-^{-1}$. 
Above $t(\pi q_1) > t(\pi q_2)$, the only contribution to $\wf'_\b( P_{\rm F}^{-1})$ comes from $ \Lambda^+$ and can be estimated using \eqref{eq:mlkfn}. In a similar vein, we can write $P_{\rm F}^{-1}= \i^{-1} \Lambda^- + P_+^{-1}$ and so the only contribution to $\wf'_\b( P_{\rm F}^{-1})$ above $t(\pi q_1) < t(\pi q_2)$ comes from $\Lambda^-$, which is estimated using \eqref{eq:mlkfn}. This way one gets the first line in \eqref{eq:wfF}. The  $P_{\aF}^{-1}$ case  is analogous.
\qed

\begin{proposition}\label{cor1} Suppose $\Lambda^\pm$ and $\tilde\Lambda^\pm$ are holographic Hadamard two-point functions. Then $\Lambda^\pm - \tilde\Lambda^\pm\in \cW^{-\infty}_{\b}(X)$.
\end{proposition}
\proof Since $\Lambda^{+}-\Lambda^{-}= \i G = \tilde\Lambda^{+}-\tilde\Lambda^{-}$, we have
\beq\label{lambdaaa}
\Lambda^{+}-\tilde\Lambda^{+}=\Lambda^{-}-\tilde\Lambda^{-}.
\eeq
The $\b$-wave front set of the LHS of \eqref{lambdaaa} is contained in $\dot\cN^+\times  \dot\cN^+$ whereas the $\b$-wave front set of the RHS is contained in $\dot\cN^-\times  \dot\cN^-$, hence the two are disjoint. Thus, both sides of \eqref{lambdaaa} have in fact empty $\b$-wave front set, and thus belong to 
$\cW^{-\infty}_{\b}(X)$.\qed

\begin{proposition}Suppose $\tilde P_+^{-1}:\Hc^{-1,-\infty}\to\Hl^{1,-\infty}$ satisfies $P \tilde P_+ =\one$, $\tilde P_+ P =\one$ and
\beq\label{eq:assfwg}
\wf'_\b(\tilde P_+^{-1})\setminus\tdiag\subset \{ (q_1,q_2) : \ q_1 \dot\sim q_2, \  t(\pi q_1)> t(\pi q_2) \}.
\eeq
Then $\tilde P_+^{-1} - P_+^{-1}\in \cW^{-\infty}_{\b}(X)$. 
\end{proposition}
\proof Suppose that $(q_1,q_2)\in\wf'_\b(\tilde P_+^{-1} - P_+^{-1})$.  Note that $P (\tilde P_+^{-1} - P_+^{-1})=0$, so $q_1,q_2\in \dot\cN$. By propagation of singularities, $(q_1',q_2)\in\wf'_\b(\tilde P_+^{-1} - P_+^{-1})$ for all $q_1'\dot\sim q_2$, in particular $(q_1',q_2)\in\wf'_\b(\tilde P_+^{-1} - P_+^{-1})$ for some $q_1'$ such that $t(\pi q_1')< t(\pi q_2)$.
But this contradicts the fact that necessarily $t(\pi q_1')\geq t(\pi q_2)$ by \eqref{eq:assfwg} and \eqref{eq:wfPpm}. This proves that the $\b$-wave front set of $\tilde P_+^{-1} - P_+^{-1}$ is empty and hence $\tilde P_+^{-1} - P_+^{-1}\in \cW^{-\infty}_{\b}(X)$.
\qeds

In a similar vein, $P_-^{-1}$, $P_\F^{-1}$ and $P_{\aF}^{-1}$ are characterized by their $\b$-wave front set uniquely modulo terms in $\cW^{-\infty}_{\b}(X)$.

\subsection{Boundary-to-boundary two-point functions} Let us recall that we defined in Subsection \ref{ss:holo} the `bulk-to-boundary' map
\[ 
\p_+ : \, x^{\nu_+}\cf\dv \to \cD'(\pX).
\]
Suppose that $\Lambda:\Hc^{-1,-\infty}(X)\to\Hl^{1,-\infty}(X)$ satisfies $P\Lambda=0$. Then by Proposition \ref{mimi}, the range $\Lambda$ is in $x^{\nu_+}\cf\dv$, and
\beq\label{eq:sdmkfsa}
\p_+ \Lambda : \Hc^{-1,-\infty}(X)\to \cD'(X)
\eeq
is continuous. Furthermore, it restricts to a continuous map 
\beq\label{eq:sdmkfsa2}
\p_+\Lambda : \Hc^{-1,\infty}(X)\to \cf(X).
\eeq
Our goal is to study the holographic data of two-point functions $\Lambda^\pm$, formally given by $\p_+ \Lambda^\pm \p_+^*$. As it is not immediately clear how to usefully define the adjoint $\p_+^*$ in the present context, instead we set for $\Lambda:\Hc^{-1,-\infty}(X)\to\Hl^{1,-\infty}(X)$ such that $P\Lambda=\Lambda P=0$,
\[
\p_+ \Lambda \p_+^*\defeq \p_+ (\p_+\Lambda^*)^*.    
\]
Since $P\Lambda^*=\Lambda^*P=0$, $\p_+\Lambda^*$ has the mapping properties as in \eqref{eq:sdmkfsa} and \eqref{eq:sdmkfsa2}, we conclude that   
\[
\p_+ \Lambda \p_+^* : \cE'(\pX) \to \cD'(\pX)
\]
is continuous.

 We now give an operatorial version of Proposition \ref{wfs}, which provides an estimate on the wave front set of $\p_+ \Lambda \p_+^*$. 

If $\Gamma\subset \be T^* X \times \be T^* X$, we denote by $\Gamma\traa{\pX\times\pX}$ the intersection $\Gamma\cap (T^* {\pX}\times T^*{\pX})$ defined by means of the embedding of  $T^*\p X$ in $\be T^*_{\pX} X$. 

\begin{proposition} Suppose $\Lambda: \Hc^{-1,-\infty}(X)\to \Hl^{1,-\infty}(X)$ is continuous and $P\Lambda=\Lambda P=0$. Then
\beq\label{eq:wfwwe}
\wf'(\p_+ \Lambda \p_+^*)\cap (T^*\pX\setminus\!\zero)\times (T^*\pX\setminus\!\zero) \subset \wf'_\b(\Lambda)\traa{\pX\times\pX}.
\eeq
\end{proposition}
\proof Suppose $(q_1,q_2)\notin \wf'_\b(\Lambda)\traa{\pX\times\pX}$, so that there exists $B_i$ elliptic at $q_i$ such that $B_1 \Lambda B_2^*\in\cW_\b^{-\infty}(X)$. By Lemma \ref{lem:comm}, there exists $B_{i,0}$ elliptic at $q_i$ and such that $\p_+ B_i = B_{i,0} \p_+$. 

Since $B_2$ preserves $x^{\nu_+}\cf\dv$, $\p_+ B_2 \Lambda^*$ is well-defined. Furthermore 
\beq\label{weflwn}
\p_+ B_2 \Lambda^*B_1^*=B_{2,0}\p_+ \Lambda^*B_1^* : \, \Hc^{-1,-\infty}(X)\to \cD'(\pX)
\eeq  
is continuous since $\p_+ \Lambda^*B_1^*$ is. Arguing exactly as in the proof of Proposition \ref{prop:wf}  we can show that \eqref{weflwn} has range in $\cf(\pX)$. It follows that its dual extends to a continuous map
\beq\label{eq:dmkkm}
B_1(\p_+ B_2 \Lambda^*)^* : \,\cE'(\pX)\to \Hl^{1,\infty}(X).
\eeq
Since $P(\p_+ B_2 \Lambda^*)^*=0$, the range of $(\p_+ B_2 \Lambda^*)^*$ is in $x^{\nu_+}\cf\dv$ and so $\p_+(\p_+ B_2 \Lambda^*)^*:  \cE'(\pX)\to\cD'(\pX)$ is well-defined. The map
\[
\p_+ B_1(\p_+ B_2 \Lambda^*)^* =B_{1,0}\p_+(\p_+ B_2 \Lambda^*)^* : \cE'(\pX)\to\cD'(\pX)
\]
is continuous (since $\p_+(\p_+ B_2 \Lambda^*)^*$ is). Furthermore, using again the argument from the proof of Proposition \ref{prop:wf} we conclude that its range is contained in $\cf(\pX)$. Since this map can also be expressed as
\[
B_{1,0}\p_+(\p_+ B_2 \Lambda^*)^*=  B_{1,0} \p_+  (\p_+\Lambda)^*  B_{2,0}^* =B_{1,0} (\p_+ \Lambda \p_+^*)  B_{2,0}^*,
\]   
we conclude that 
\[
 B_{1,0} (\p_+ \Lambda \p_+^*)  B_{2,0}^*  : \, \cE'(X)\to\cf(X).
\]
This  shows that $(q_1,q_2)\notin\wf(\p_+ \Lambda \p_+^*)$.
\qeds

Note that because of how we defined $\wf'_\b$, the estimate \eqref{eq:wfwwe} gives no information about possible singularities in $\zero \times (T^*\pX\setminus\!\zero)$ or $(T^*\pX\setminus\!\zero) \times \zero$. In practice however these can often be ruled out otherwise, as illustrated in the result below.

\begin{theorem}\label{hololo} Suppose $(X,g)$ is an asymptotically $\AdS$ spacetime and assume $\nu>0$. If $\Lambda^\pm$ is a pair of holographic Hadamard two-point functions then
\beq\label{betterwf}
\wf'(\p_+ \Lambda^\pm \p_+^*)\subset\wf'_\b(\Lambda^\pm)\traa{\pX\times\pX}\subset (\dot\cN^\pm\times\dot\cN^\pm)\traa{\pX\times\pX}.
\eeq
Furthermore, if $\tilde\Lambda^\pm$ is another pair of holographic Hadamard two-point functions then $\tilde\Lambda^\pm-\Lambda^\pm$ has smooth Schwartz kernel.
\end{theorem}
\proof In order to conclude \eqref{betterwf} from \eqref{eq:wfwwe} and the definition of holographic Hadamard two-point functions, it suffices to prove that 
\[
\wf'(\p_+ \Lambda^\pm \p_+^*)\subset (T^*\pX\setminus\!\zero)\times (T^*\pX\setminus\!\zero).
\] 
This is easily shown using the positivity of $\p_+ \Lambda^\pm \p_+^*$ in a similar vein as in Lemma \ref{lempos}, we refer to \cite{radzikowski2} or the proof of \cite[Prop. 3.1]{VW} for the precise argument. 

The second statement is proved analogously to Proposition \ref{cor1}.\qeds

We can rephrase \eqref{betterwf} in a slightly more explicit way using coordinates $(x,y)$ on a neighborhood $U$ of a point on $\pX$ as before, with $y=(y_0,\dots,y_{n-2})$ coordinates on $\p X$.

The assumptions on the metric $g$ (Definition \ref{defads}) imply that the restriction of the principal symbol of $\Box_{\tilde g}$ to the boundary is of the form
\[
\tilde p (0,y,\xi,\zeta)=-\xi^2 +\zeta \cdot h^{-1}(y) \zeta,
\]
where $h$ is a Lorentzian metric on $\p X$. Thus, locally over the boundary, the compressed characteristic set $\dot\cN$ is
\[
\dot\cN\cap \be T^*_{\pX\cap U} X = \{ (0,y,0,\zeta) :  \  \zeta \cdot h^{-1}(y)\zeta \geq 0, \ \zeta\neq 0 \}.
\]  
The coordinates can be further adjusted in such way that the sign of $\zeta_0$ distinguishes between $\dot\cN^+$ and $\dot\cN^-$. With these choices, \eqref{betterwf} states that
\beq\label{eq:finalwf}
\wf'(\p_+ \Lambda^\pm \p_+^*)\cap T^*_U\pX\subset  \cN_U^\pm \times \cN_U^\pm,
\eeq
where $\cN_U^\pm=\{ (y,\zeta)\in T^*_U\pX :  \  \zeta \cdot h^{-1}(y)\zeta \geq 0, \ \pm{\zeta_0}>0 \}$. This estimate can be improved using Theorem \ref{prop:wfs} to account for the fact that $q_1$ is connected with  $q_2$ by a generalized broken bicharacteristic  if $(q_1,q_2)\in \wf'_\b(\Lambda^\pm)$. 

Let us point out that the estimate \eqref{eq:finalwf} allows for a larger wave front set than that of Hadamard two-point functions on a  globally hyperbolic spacetime. However, it is still the case that $\wf'(\p_+ \Lambda^\pm \p_+^*)\subset \pm(\Gamma\times\Gamma)$ for some $\Gamma\subset T^*\pX \setminus\zero$ such that $\Gamma \cap -\Gamma =\{0\}$ (where the minus sign means replacing $(y,y',\zeta,\zeta')$ by $(y,y',-\zeta,-\zeta')$, and similarly for  $T^*\pX$), which is the basic property used in the perturbative construction of interacting fields \cite{BF00}.

\appendix
\section{}
\subsection{The $\b$-calculus}\label{appb}

In this appendix we briefly recall basic material on the $\b$-pseudodifferential calculus, following mainly \cite{corners,vasy}. Other useful references include \cite{melrose88,edges,hintz,GHV}, cf.~\cite{melrose,hoermander} for textbook accounts.

As in the main part of the text, $X$ is an $n$-dimensional manifold with boundary  $\p X$. We will use here various definitions from Subsections \ref{ss:b} and \ref{ss:conormal}.

We denote by $S^s(\be T^*X)$ the set of symbols of order $s$ on $\be T^*X$ (defined as for any vector bundle over $X$), and by $S^s_{\rm ph}(\be T^*X)$ the subset of (fiberwise) poly-homogeneous ones. 

We will consider here only the one-step poly-homogeneous $\b$-pseudodifferential operator classes. Namely,
\[
\Psi^s_\b(X)\defeq \Op\big(S^s_{\rm ph}(\be T^*X)\big)+\Psi_\b^{-\infty}(X),
\] 
where $\Op$ is a suitable quantization map (given below) and $\Psi_\b^{-\infty}(X)$ is the ideal of regularizing $\b\Psi$DOs (see \cite{melrose} for its precise definition, here we will only use the fact that $\Psi^{-\infty}_\b(X)=\bigcap_{s\in\rr}\Psi^s_\b(X)$ and that $\Psi_\b^{-\infty}(X)$ maps continuously $\Hc^{k,-\infty}(X)$ to $\Hl^{k',\infty}(X)$ for all $k,k'\in\zz$). Over a local coordinate chart $U$ with coordinates $(x,y)$ (where as usual $x$ is a boundary-defining function), if $a\in S^s(\be T^*X)$ is supported in $\be T_K^*X$ with $K\subset U$ compact, $\Op(a)$ can be defined by the oscillatory integral
\[
\bea
\Op(a)u(x,y)=(2\pi)^{-n}\int &\e^{\i((x-x'){\xi}+(y-y')\cdot\zeta)}\phi\big(\textstyle\frac{x-x'}{x}\big) \\ &\,\times a(x,y,x\xi,\zeta) u(x',y') dx'\,dy'\,d\xi\,d\zeta,
\eea
\]
where the integral in $x'$ is over $[0,\infty)$ and $\phi\in\cf_\c((-1/2,1/2))$ is identically $1$ near $0$. This definition is then made into a global one using a partition of unity in the usual way. Then in particular $\Diff_\b^s(X)\subset \Psi_\b^s(X)$.

 Recall that $\cf(X)$ is the space of smooth functions on $X$ in the sense of extendability across the boundary, and $\dot\cC^{\infty}(X)$ is the space of smooth functions on $X$ vanishing with all derivatives at the boundary $\p X$. A standard fact says that if $A\in\Psi_\b^s(X)$ has properly supported Schwartz kernel then it maps continuously $A:\cfd(X)\to\cfd(X)$ and $A:\cf(X)\to\cf(X)$ (and therefore such pseudo-differential operators can be composed). Throughout the text we assume that all the $\b$-pseudo\-differential operators that we consider have this property: this can always be ensured by appropriate cutoffs, which play no essential r\^ole here as $\b$-pseudo\-differential operators appear only as a device to microlocalize in $\be T^*X$. With this assumption, if $k\in\zz$ and $s\in\rr$ one can show that $A\in\Psi_\b^0(X)$ extends to a continuous map
\beq
A : \Hb^{k,s}(X)\to\Hb^{k,s}(X),
\eeq
see \cite[Lem.~5.8]{vasy} for the proof and a more precise statement on the operator norm (recall that the spaces $\Hb^{k,s}(X)$ were defined in \eqref{eq:defks} in the main part of the text).

 There is also a principal symbol map $\sigma_{\b,s}:\Psi_\b^s(X)\to \cf(\be T^*X)$ with values in homogeneous functions of degree $s$, such that $\sigma_{\b,s}\big(\Psi_\b^{s-1}(X)\big)=\{0\}$.    

One denotes
\[
\Psi_\b(X)=\textstyle\bigcup_{s\in\rr} \Psi^s_\b(X),
\]
and similarly $S^{-\infty}(\be T^*X)=\textstyle\bigcap_{s\in\rr} S^s(\be T^*X)$ for order $-\infty$ symbols.

Let us denote by $A^*$ the adjoint of $A\in\Psi^s_\b(X)$ with respect to the $L^2(X,g)$-inner product, defined using an arbitrary smooth pseudo-Riemannian metric $g$. Then $A^*\in\Psi^s_\b(X)$ and $\sigma_{\b,s}(A^*)=\overline{\sigma_{\b,s}(A)}$. Furthermore, $\Psi_\b(X)$ has the structure of a filtered algebra in the sense that if $A_i\in\Psi^{s_i}_\b(X)$, $i=1,2$, then $A_1 A_2\in \Psi^{s_1+s_2}_\b(X)$. Moreover,
\[
\sigma_{\b,s_1+s_2}(A_1 A_2)=\sigma_{\b,s_1}(A_1)\sigma_{\b,s_2}(A_2),
\]
and so $[A_1,A_2]\in\Psi^{s_1+s_2-1}_\b(X)$. A less obvious fact (efficiently proved using the so-called normal operator family, see e.g.~\cite{corners}) is that if $A\in\Psi_\b^s(X)$ then the commutator $[x D_x,A]$ belongs to $x \Psi_\b^s(X)$ rather than merely to $\Psi_\b^s(X)$. Another useful feature of the $\b$-calculus is that if $l\in \rr$ and $A\in \Psi^s_\b(X)$, then 
\[
x^{-l}A x^l\in \Psi^s_\b(X) \mbox{ and }  \sigma_s(x^{-l}A x^l)=\sigma_s(A).  
\]
A consequence of this is that any $A\in\Psi^s_\b(X)$ maps $x \cf(X)$ to itself, and so $(A u)\tra{\pX}$ depends only on $u\tra{\pX}$. Thus, $\b$-pseudo\-differential operators preserve Dirichlet boundary conditions.

Similarly as in the pseudodifferential calculus on manifolds without boundary, there is an operator $\b$-wave front set\footnote{It is usually denoted by $\wf'_\b(A)$ in the literature, here however the notation $\wf'_\b(A)$ is reserved for the more general operator $\b$-wave front set defined in Section \ref{sec:propagators}.} $\wf^\Psi_\b(A)\subset \be T^* X$ that indicates where in `phase space' a given pseudo\-differential operator $A\in\Psi_\b(X)$ is not in $\Psi_\b^{-\infty}(X)$.

\begin{definition} For $A\in\Psi_\b(X)$, $q\in \be T^*X\setminus \zero$ is \emph{not} in $\wf^\Psi_\b(A)$ if $q$ has a conic neighborhood on which $a$ is the restriction of a symbol in $S^{-\infty}(\be T^*X)$.     
\end{definition}

This means in particular that $\wf^\Psi_\b(A)$ is empty if and only if $A\in\Psi^{-\infty}_\b(X)$. Furthermore, the operator $\b$-wave front set defined in this way satisfies
\[
\wf^\Psi_\b(A_1 A_2)\subset \wf^\Psi_\b(A_1)\cap \wf^\Psi_\b(A_2).
\]
One says that $A\in\Psi^s_\b(X)$ is \emph{elliptic} (at $q\in\be T^* X\setminus\zero$, resp. on $K\subset \be T^* X\setminus\zero$) if $\sigma_{\b,s}(A)$ is invertible (at $q$, resp. on $K$). If $A\in\Psi^s_\b(X)$ is elliptic then there exists $A^{\inv}\in\Psi_\b^{-s}(X)$ such that $A A^\inv-\one,A^\inv A -\one \in \Psi_\b^{-\infty}(X)$. More generally, suppose that $K\subset \be S^*X$ is compact and $A\in\Psi^s_\b(X)$ is elliptic on $K$. Then there exists a \emph{microlocal parametrix}, i.e.~an operator $A^{\inv}\in\Psi_\b^{-s}(X)$ that satisfies
\[
\wf^\Psi_\b(A^\inv A - \one)\cap K = \emptyset, \ \ \wf^\Psi_\b( A A^\inv - \one)\cap K = \emptyset.
\]
Thus in particular, if $A$ is elliptic at $q\in\be T^* X\setminus\zero$ then there exists $A^{\inv}\in\Psi_\b^{-s}(X)$ such that $q\notin\wf^\Psi_\b(A^\inv A - \one)$ and $q\notin \wf^\Psi_\b(A A^\inv - \one)$.

\medskip

\subsection*{Acknowledgments} The author is deeply grateful to Andr\'as Vasy for all the discussions and helpful comments. The author would also like to thank Claudio Dappiaggi, Christian G\'erard, Peter Hintz, Jacques Smulevici and Jochen Zahn for stimulating discussions and useful remarks. Financial support from the ANR-16-CE40-0012-01 grant is gratefully acknowledged. The author is also grateful to the Erwin Schr\"odinger Institute in Vienna for its hospitality during the program ``Modern theory of wave equations''.


\begin{thebibliography}{plain}

\bibitem[AIS78]{isham} S. J. Avis, C. J. Isham, and D. Storey, {\em Quantum field theory in Anti-de Sitter space-time}, Phys.
Rev. D, 18:3565-3576, (1978).

\bibitem[An04]{anderson} M. T. Anderson, {\em On the structure of asymptotically de Sitter and anti-de Sitter spaces}, Adv. Theor. Math. Phys., 8(5), (2004), 861--894.

\bibitem[Ba08]{bachelot3} A. Bachelot, {\em The Dirac System on the Anti-de Sitter Universe}, Comm. Math. Phys. 283 (2008) 127-167.

\bibitem[Ba11]{bachelot4} A. Bachelot, {\em The Klein--Gordon equation in the Anti-de Sitter cosmology}, Journal de math\'ematiques pures et appliqu\'es, 96(6), 527--554 (2011)

\bibitem[Ba13]{bachelot2} A. Bachelot, {\em New dynamics in the Anti-de Sitter universe ${AdS}^5$}. Comm. Math. Phys., 320(3), 723-759 (2013)

\bibitem[Ba16]{bachelot} A. Bachelot, {\em On the Klein-Gordon equation near a De Sitter brane in an Anti-de Sitter bulk}, 105 (2), (2016), 165-197.

\bibitem[BD15]{benini} M. Benini, C. Dappiaggi, {\em Models of free quantum field theories on curved backgrounds}, in: Advances in
Algebraic Quantum Field Theory, Springer (2015).


\bibitem[BGP07]{BGP07} C. B\"ar, N. Ginoux, F. Pf\"affle, {\em Wave equation on Lorentzian
manifolds and quantization}, ESI Lectures in Mathematics and Physics, EMS (2007).

\bibitem[BEM02]{BEM02} J. Bros, H. Epstein, U. Moschella, {\em Towards a general theory of quantized fields on the
anti-de Sitter space-time}, Commun. Math. Phys. 231 (2002), 481.

\bibitem[BF00]{BF00} R. Brunetti, K. Fredenhagen, {\em Microlocal Analysis and Interacting Quantum Field Theories: Renormalization
	on Physical Backgrounds}, Comm. Math. Phys. {\bf 208} (2000), 623-661.

\bibitem[BF14]{brumfredenhagen} M. Brum, K. Fredenhagen, {\em `Vacuum-like' Hadamard states for quantum fields on curved spacetimes}, Class. Quantum Grav. 31 , no. 2, (2014), 025024.

\bibitem[BF82]{breiten} P. Breitenlohner, D.Z. Freedman, {\em Positive energy in anti-de Sitter backgrounds and gauged extended supergravity} Phys. Lett. B, 115(3):197-201, (1982).


\bibitem[BFQ16]{BFQ} A. Belokogne, A. Folacci, J. Queva, {\em Stueckelberg massive electromagnetism in de Sitter and anti-de Sitter spacetimes: Two-point functions and renormalized stress-energy tensors}, Phys. Rev. D 94, (2016), 105028.

\bibitem[BJ14]{BJ} M. Brum, S. E. Jor\'as, {\em Hadamard state in Schwarzschild-de Sitter spacetime}, Class. Quantum Grav. 32, no. 1 (2014).

\bibitem[CG14]{CG} S. Curry, R. Gover, {\em An introduction to conformal geometry and tractor calculus, with a view to applications in General Relativity}, preprint \texttt{arXiv:1412.7559} (2014)

\bibitem[Da13]{dang} N.V. Dang, {\em Renormalization of quantum field theory on curved space\-times, a causal approach}, Ph.D. thesis, Paris Diderot University, (2013).

\bibitem[DF16]{DF} C. Dappiaggi, H. R. C. Ferreira, {\em Hadamard states for a scalar field in anti-de Sitter spacetime with arbitrary boundary conditions}, Phys. Rev. D 94 (2016), 125016.


\bibitem[Do17]{dold} D. Dold, {\em Unstable mode solutions to the Klein-Gordon equation in Kerr-anti-de Sitter spacetimes},  Commun. Math. Phys. 350, 2 (2017), 639--697.
 
\bibitem[DNP16]{casimir} C. Dappiaggi, G. Nosari, N. Pinamonti, {\em The Casimir effect from the point of view of algebraic quantum field theory},  Math. Phys. Anal. Geom. 19 (2016).

\bibitem[DG13]{derger} J. Derezi\'nski, C. G\'erard, {\em Mathematics of Quantization and Quantum Fields}, Cambridge Monographs in Mathematical Physics, Cambridge University Press (2013).


\bibitem[DH72]{DH72} J.J. Duistermaat, L. H\"{o}rmander, \textsl{Fourier integral
operators II}, Acta Math. {\bf 128}  (1972), 183--269.

\bibitem[DR02]{DR1} M. D\"utsch, K.-H. Rehren, {\em A comment on the dual field in the AdS-CFT correspondence},
Lett. Math. Phys. 62 (2002) 171--184.

\bibitem[DR03]{DR2} M. D\"utsch, K.-H. Rehren, {\em Generalized free fields and the AdS-CFT correspondence}, Ann.
Henri Poincar´e 4 (2003) 613--635.

\bibitem[DR11]{DR3} M. D\"utsch, K.-H. Rehren, {\em Protecting the conformal symmetry via bulk renormalization on Anti de Sitter space}, Comm. Math. Phys. (2011), 307:315.

\bibitem[DR16]{DR} J. Derezi\'nski, S. Richard, {\em On Schr\"odinger operators with inverse
square potentials on the half-line}, Ann. Henri Poincar\'e (2016), DOI
10.1007/s00023-016-0520-7.


\bibitem[EK15]{kamran} A. Enciso, N. Kamran, {\em A singular initial-boundary value problem for nonlinear wave equations
and holography in asymptotically anti-de Sitter spaces}, J. Math. Pures Appl. 103 (2015), 1053--1091.

\bibitem[FMR16]{FMR}  F. Finster, S. Murro, C. R\"oken, {\em The fermionic projector in a time-dependent external potential:
Mass oscillation property and Hadamard states}, J. Math. Phys. 57 (2016), 072303.

\bibitem[FNW81]{FNW} S.A. Fulling, F.J. Narcowich, R.M. Wald, {\em Singularity structure of the two-point function in quantum field theory in curved space-time, II}, Annals of Physics, {\bf 136} (1981), 243-272. 

                                                                                                                                          
\bibitem[FR15]{FR15} K. Fredenhagen, K. Rejzner, {\emph Perturbative algebraic quantum field theory}, in: Mathematical Aspects of Quantum Field Theories, Springer, (2015), 17--55.

\bibitem[FR16]{FR16} K. Fredenhagen, K. Rejzner, {\emph Quantum field theory on curved spacetimes: Axiomatic framework and examples}, Journal of Mathematical Physics 57(3), (2016), 031101.

\bibitem[FSW78]{FSW} S.A. Fulling, M. Sweeny, R.M. Wald, {\em Singularity structure of the two-point function in quantum field theory in curved space-time}, Comm. Math. Phys. {\bf 63} (1978), 257-264.

\bibitem[FV13]{FV} C.J. Fewster, R. Verch, {\em The necessity of the Hadamard condition}, Class. Quant. Grav. 30, 235027 (2013).

\bibitem[FV15]{FV2} C.J. Fewster, R. Verch, {\em Algebraic quantum field theory in curved spacetimes}, in: Advances in Algebraic Quantum Field Theory, Springer (2015). 

																																																																	\bibitem[Ga15]{gannot} O. Gannot, {\em Elliptic boundary value problems for Bessel operators, with applications to Anti-de Sitter spacetimes}, preprint \texttt{arxiv:1507.02794} (2015)

\bibitem[Ga16]{gannot2} O. Gannot, {\em Existence of quasinormal modes for Kerr-AdS black holes}, acc. in Annales Henri Poincar\'e, \texttt{arXiv:1602.08147}, (2016).

																																																																																					

\bibitem[GHV16]{GHV} J. Gell-Redman, N. Haber, A. Vasy, \textsl{The Feynman propagator on perturbations of Minkowski space}, Comm. Math. Phys. 342 (1), (2016), 333--384.
 
\bibitem[GL91]{grahamlee} C.R. Graham, J.M. Lee, \textsl{Einstein metrics with prescribed conformal infinity on the ball},
Adv. Math., 87 (2), (1991), 186--225.

\bibitem[GLW15]{GLW} R. Gover, E. Latini, A. Waldron, {\em Poincar\'e--Einstein holography for forms via conformal geometry in the bulk}. Mem. Amer. Math. Soc. 235 (2015), 1106.  

\bibitem[GOW17]{GOW} C. G\'erard, O. Oulghazi, M. Wrochna, \textsl{Hadamard states for the Klein-Gordon equation on Lorentzian manifolds of bounded geometry}, Commun. Math. Phys. 352 (2), (2017), 352--519..


\bibitem[GW14a]{gover} R. Gover, A. Waldron, {\em Boundary calculus for conformally compact manifolds}, Indiana Univ. Math. J. 63 (2014), no. 1, 119--163. 

\bibitem[GW14b]{GW} C. G\'erard, M. Wrochna, \textsl{Construction of Hadamard states by pseudo-differential calculus}, Comm. Math. Phys. \textbf{325} (2) (2014), 713--755.

\bibitem[Hi15]{hintz} P. Hintz, {\em Global analysis of linear and nonlinear wave equations on cosmological spacetimes}, PhD thesis, Stanford University (2015)

\bibitem[HLSW15]{hlsw} G. Holzegel, J. Luk, J. Smulevici, C. Warnick, {\em Asymptotic properties of linear field equations in anti-de Sitter space}, \texttt{arXiv:1502.04965} (2015).


 \bibitem[Ho01]{hollands} S. Hollands, \textsl{{The Hadamard Condition for Dirac Fields and Adiabatic
  States on Robertson-Walker space\-times}}, \newblock Comm. Math. Phys. \textbf{ 216} (2001), 635--661. 

\bibitem[Ho12]{holzegel} G. Holzegel {\em Well-posedness for the massive wave equation on asymptotically anti-de sitter space-
times}, Journal of Hyperbolic Differential Equations, 09(02):239-261, (2012).


\bibitem[H\"o07]{hoermander} L. H\"ormander, {\em The analysis of linear partial differential operators I-IV}, Classics
in Mathematics, Springer (2007).

\bibitem[HS14]{HS} G. Holzegel, J. Smulevici, {\em Quasimodes and a Lower Bound on the Uniform Energy Decay Rate for Kerr-AdS Spacetimes}, Anal. PDE 7, No. 5, 1057-1090 (2014).

\bibitem[HV16]{HV}  P. Hintz, A. Vasy, {\em The global non-linear stability of the Kerr-de Sitter family of black holes}, preprint \texttt{arXiv:1606.04014}, (2016).

\bibitem[HW02]{HW1} S. Hollands, R.M. Wald, {\em Existence of local covariant time ordered products of quantum fields in curved spacetime}, Comm. Math. Phys. 231, no. 2 (2002), 309--345.

\bibitem[HW05]{HW2} S. Hollands, R.M. Wald, {\em Conservation of the stress tensor in perturbative
interacting quantum field theory in curved spacetimes}, Rev. Math. Phys. 17.3 (2005), 277-311.

\bibitem[HW14]{holzegelwarnick} G. Holzegel, C.M. Warnick, {\em Boundedness and growth for the massive wave equation on
asymptotically anti-de Sitter black holes}, J. Funct. Anal. 266 (4)15, (2014), 2436-2485.

\bibitem[HW15]{HW} S. Hollands, R.M. Wald, {\em Quantum fields in curved spacetime}, in: General Relativity and Gravitation: A Centennial Perspective, Cambridge University Press (2015).

\bibitem[I-R14]{guillaume} G. Idelon--Riton, {\em Scattering theory for the Dirac equation in Schwarzschild-Anti-de Sitter space-time}, preprint \texttt{arXiv:1412.0869} (2014).


\bibitem[IW02]{ishibashiwald0} A. Ishibashi, R.M. Wald, {\em Dynamics in non-globally-hyperbolic static spacetimes: 2. General analysis of prescriptions for dynamics}, Class. Quant. Grav. 20, 16, (2004), 3815--3826.

\bibitem[IW04]{ishibashiwald} A. Ishibashi, R.M. Wald, {\em Dynamics in nonglobally hyperbolic static space-times: 3. Anti-de
Sitter space-time}, Class. Quant. Grav., 21, (2004), 2981-3014.

\bibitem[Ju96]{junker} W. Junker, {\em Hadamard States, adiabatic vacua and the construction of physical states for scalar quantum fields on curved space\-time}, Rev. Math. Phys. {\bf 8}, (1996), 1091--1159.



\bibitem[Ka92]{Kay} B.S. Kay, {\em The principle of locality and quantum field theory on (non globally
hyperbolic) curved spacetimes}, Rev. Math. Phys. (Special Issue), (1992), 167--195.



\bibitem[KL08]{KL} B.S. Kay, P. Larkin, {\em Pre-Holography}, Phys. Rev. D 77  (2008), 121501R.

\bibitem[KM15]{KM} I. Khavkine, V. Moretti, {\em Algebraic QFT in curved spacetime and quasifree Hadamard states: an introduction}, in: Advances in Algebraic Quantum Field Theory, Springer (2015).

\bibitem[KO14]{KO} B.S. Kay,  L. Ort{\'\dotlessi}z,  {\em Brick walls and AdS/CFT}, Gen. Relativ. Gravit. 46 (2014), 1727.

\bibitem[KT14]{truc} H. Kova\v{r}{\'\dotlessi}k, F. Truc, {\em Schr\"odinger operators on a half-line with inverse square potentials}, Math.
Model. Nat. Phenom. 9 (2014), 170-176.

\bibitem[KW91]{KW} Kay, B.S., Wald, R.M.: {\em Theorems on the uniqueness and thermal properties of stationary,
nonsingular, quasifree states on spacetimes with a bifurcate Killing horizon}, Phys. Rep.
207, 49 (1991)

\bibitem[KW15]{KW15} C. Kent, E. Winstanley, {\em Hadamard renormalized scalar field theory on anti-de Sitter spacetime}, Physical Review D 91.4 (2015), 044044.

\bibitem[Le97]{lebeau} G. Lebeau, {\em Propagation des ondes dans les vari\'et\'es \`a coin}, Ann. Scient. \'Ec. Norm. Sup. 30, (1997), 429--497.

\bibitem[Ma99]{maldacena} J. Maldacena, {\em The large-N limit of superconformal field theories and supergravity}, Int. J. Theor. Phys. 38.4 (1999), 1113-1133.

\bibitem[Me88]{melrose88} R. Melrose, \textsl{Transformation of boundary problems}, Acta Math., 147(3--4), (1993), 149--236.

\bibitem[Me93]{melrose} R. Melrose, \textsl{The Atiyah-Patodi-Singer index theorem}, Vol. 4. Wellesley: AK Peters, (1993).

\bibitem[MM87]{MM} R. Mazzeo, R. Melrose, \textsl{Meromorphic extension of the resolvent on complete spaces with asymptotically constant negative curvature}, J. Func. Anal. 75 (1987), 260--310.

\bibitem[MVW08]{MVW} R. Melrose, A. Vasy, J. Wunsch, \textsl{Propagation of singularities for the wave equation on manifolds with edges}, J. Func. Anal. 75 (1987), 260--310.


\bibitem[Mo14]{morrison} I.A. Morrison {\em Boundary-to-bulk maps for AdS causal wedges and the Reeh-Schlieder property in holography}, J-HEP. (5):1-29 (2014).

\bibitem[Ra96a]{radzikowski} M. Radzikowski, {\em Micro-local approach to the Hadamard condition in quantum field theory on curved space-time}, Comm. Math. Phys. {\bf 179} (1996), 529--553.

\bibitem[Ra96b]{radzikowski2} M. Radzikowski, {\em A Local to global singularity theorem for quantum field theory on curved spacetime},
Comm. Math. Phys. 180, 1 (1996).




\bibitem[Re00a]{rehren} K.-H. Rehren, {\em Local quantum observables in the anti-de Sitter-conformal QFT correspondence}, Phys. Lett. B 493.3 (2000), 383-388.

\bibitem[Re00b]{rehren2} K.-H. Rehren, {\em Algebraic holography}, Ann. Henri Poincar\'e 1, (2000), 607.

\bibitem[Ri07a]{ribeiro1} P. L. Ribeiro,  {\em Algebraic Holography in Asymptotically Simple, Asymptotically AdS SpaceTimes}, Progress in Mathematics 251, (2007), 253--270.

\bibitem[Ri07b]{ribeiro2} P. L. Ribeiro, {\em Structural and Dynamical Aspects of the AdS-CFT Correspondence: a Rigorous
Approach}, Ph.D. Thesis, University of S\~ao Paulo (2007)

\bibitem[Sa10]{sanders2} K. Sanders,  {\em Equivalence of the (Generalised) Hadamard and Microlocal Spectrum Condition for (Generalised) Free Fields in Curved Spacetime}, Commun. Math. Phys. 295, 2 (2010), 485--501.

\bibitem[Sa15]{sanders} K. Sanders,  {\em On the Construction of Hartle-Hawking-Israel States Across a
Static Bifurcate Killing Horizon}, Lett. Math. Phys. 105, 4 (2015), 575--640.

\bibitem[S\'a05]{sanchezstatic} M. S\'anchez, {\em On the geometry of static spacetimes}, Nonlinear Analysis, 63 (2005), 455--463.

\bibitem[SV01]{SV} H. Sahlmann, R. Verch, {\em Microlocal spectrum condition and Hadamard form for vector-valued quantum fields in curved spacetime}, Rev. Math. Phys., 13(10) (2001), 1203-1246.

\bibitem[Va08a]{corners} A. Vasy, \textsl{Propagation of singularities for the wave equation on manifolds with corners}, Ann. of Math. (2) 168(3), (2008), 749--812.

\bibitem[Va08b]{edges} A. Vasy, \textsl{Diffraction by edges}, Modern Phys. Lett. B, 22, (2008), 2287--2328.

\bibitem[Va10]{desitter} A. Vasy, \textsl{The wave equation on asymptotically de Sitter-like spaces}, Adv. Math., 223(1), (2010), 49--97.


\bibitem[Va12]{vasy} A. Vasy, {\em The wave equation on asymptotically Anti-de Sitter spaces}, Analysis \& PDE, 5, (2012), 81--144.
 
\bibitem[Va16]{positive} A. Vasy, \textsl{On the positivity of propagator differences}, Ann. Henri Poincar\'e, DOI 10.1007/s00023-016-0527-0 (2016). 

\bibitem[VW15]{VW} A. Vasy, M. Wrochna \textsl{Quantum fields from global propagators on asymptotically
Minkowski and extended de Sitter spacetimes}, preprint \texttt{arXiv:1512.08052}, (2015).

\bibitem[Wa80]{wald1} R.M. Wald, {\em Dynamics in nonglobally hyperbolic, static space‐times}, J. Math. Phys., 21, 2802-2805 (1980).

\bibitem[Wa13]{warnick1} C.M. Warnick, {\em The Massive Wave Equation in Asymptotically AdS Spacetimes}, Comm. Math. Phys., 321(1):85-111, (2013).

\bibitem[Wa15]{warnick2} C.M. Warnick, {\em On quasinormal modes of asymptotically Anti-de Sitter black holes},  Comm. Math. Phys., 333(2):959-1035, (2015).

\bibitem[Za15]{zahn} J. Zahn, {\em Generalized Wentzell boundary conditions and holography}, preprint \texttt{arXiv:1512.05512} (2015).
 
\bibitem[YG09]{yagdjian} K. Yagdjian, A. Galstian, {\em The Klein-Gordon equation in Anti-de Sitter spacetime}, Rend. Sem. Mat. Univ. Pol. Torino, 67:291-292, (2009).



\end{thebibliography}
\end{document}